\documentclass[12pt]{article}
\usepackage{latexsym}
\usepackage{amsmath}
\usepackage{amssymb}
\usepackage{epsfig,graphics}
\usepackage{graphicx}
\usepackage{booktabs}
\usepackage{multirow}
\usepackage{caption}
\usepackage{subcaption}
\usepackage{tikz}

\newcommand{\be}{\begin{equation}}
\newcommand{\ee}{\end{equation}}

\begin{document}

\begin{titlepage}

\vspace*{0.6in}
 
\begin{center}
{\large\bf Closed flux tubes in $D=2+1$ $SU(N)$ gauge theories: \\
dynamics  and  effective string description.}\\
\vspace*{0.75in}
{Andreas Athenodorou$^{a,b}$ and Michael Teper$^{c}$\\
\vspace*{.25in}
$^{a}$Department of Physics, University of Cyprus, POB 20537, 1678 Nicosia, Cyprus\\
\vspace*{.1in}
$^{b}$Computation-based Science and Technology Research Center, The Cyprus Institute, 20 Kavafi Str., Nicosia 2121, Cyprus \\
\vspace*{.1in}
$^{c}$Rudolf Peierls Centre for Theoretical Physics, University of Oxford,\\
1 Keble Road, Oxford OX1 3NP, UK}
\end{center}

\vspace*{0.4in}

\begin{center}
{\bf Abstract}
\end{center}

We extend our earlier calculations of the spectrum of closed flux tubes 
in $SU(N)$ gauge theories in $2+1$ dimensions, with a focus on 
questions raised by recent theoretical progress on the effective
string action of long flux tubes and the world-sheet
action for flux tubes of moderate lengths. Our new calculations in
$SU(4)$ and $SU(8)$ provide evidence that the leading $O(1/l^\gamma)$
non-universal correction to the flux tube ground state energy does indeed
have a power ${\gamma \geq 7}$. We perform a study in $SU(2)$, where
we can traverse the length at which the Nambu-Goto ground state becomes 
tachyonic, to obtain an all-$N$ view of the spectrum.
Our comparison of the $k=2$ flux tube excitation energies in $SU(4)$ 
and $SU(6)$ suggests that the massive world sheet excitation associated 
with the $k=2$ binding has a scale that knows about the group and hence 
the theory in the bulk, and we  comment on the potential implications of world 
sheet massive modes for the bulk spectrum. We provide a quantitative 
analysis of the surprising (near-)orthogonality of flux tubes carrying 
flux in different $SU(N)$ representations, which implies that their 
screening by gluons is highly suppressed even at small $N$.  

\vspace*{0.95in}

\leftline{{\it E-mail:} athenodorou.andreas@ucy.ac.cy, mike.teper@physics.ox.ac.uk}

\end{titlepage}

\setcounter{page}{1}
\newpage
\pagestyle{plain}

\tableofcontents

\section{Introduction}
\label{section_intro}

In this paper we extend our earlier calculations on the spectrum
of closed confining flux tubes that wind around a spatial torus
\cite{AABBMT11,AAMT13,AABBMT}
in $SU(N)$ gauge theories in $2+1$ dimensions.
Our main motivation is to address questions raised by the remarkable recent
analytic progress both in understanding the universal physics of long flux
tubes 
\cite{OAZK,OA}
(see also
\cite{SD_long})
and in understanding the simplicity of the spectrum of shorter flux tubes
\cite{SD_short,SD_massive}.
In addition we will emphasise a further feature that is surprising and
potentially of significant theoretical interest. 

Our earlier calculations of the $l$-dependence of the flux tube spectrum
\cite{AABBMT11,AAMT13}
were primarily in $SU(6)$. In this paper we extend those calculations, as
well as providing new calculations in $SU(2),\,SU(4)$ and $SU(8)$. This
enables us to say more about the $N$-dependence of the spectra and, in the
case of $SU(2)$, to see what happens when $l$ is decreased past the value
at which the Nambu-Goto ground state becomes tachyonic.

We begin by sketching, in the next Section, the theoretical background. We
then outline, in Section~\ref{section_fluxtubes}, the most relevant details
of the lattice calculations; a fuller description may be found in
\cite{AABBMT11,AAMT13}.
In Section~\ref{section_nonuniversal} we test the predictions of
\cite{OAZK,OA}
for the leading non-universal power correction to `long' flux tubes.
In Section~\ref{section_massive} we examine more closely the resonant 
massive mode tentatively discussed in our earlier work on $k$-strings
\cite{AAMT13}
and then unambigously identified
\cite{SD_massive}
using the powerful formalism based on the Thermodynamic Bethe Ansatz.
In  Section~\ref{section_overlaps} we probe more quantitatively the
remarkable absence of screening of all but the shortest flux tubes when
these carry flux in higher representations, with explicit examples
for $k=2$ flux in $SU(4)$ and adjoint flux in both $SU(2)$ and $SU(4)$.
In  Section~\ref{section_spectra} we compare the $l$-dependence of the
spectra of fundamental flux tubes in $SU(2),\,SU(4),\,SU(6)$ and $SU(8)$.
We also analyse adjoint flux tubes in $SU(2)$ and $SU(4)$, and critically
examine the remarkable agreement with Nambu-Goto of flux tubes with non-zero
momentum along the flux tube. A summary and discussion of our results
is given in Section~\ref{section_conclusion}.

Most of the qualitative discussion in this paper applies to flux tubes in 
$D=3+1$ as much as to $D=2+1$. However the specific results do
differ and everything we say in this paper should be taken to be for
$D=2+1$ unless explicitly stated otherwise.

\section{Some background}
\label{section_background} 

We begin with a brief overview of the `Nambu-Goto spectrum' since it
has been found to provide a good approximation to the observed spectrum
of fundamental flux tubes
\cite{AABBMT11,AAMT13,AABBMT}.
(See also
\cite{caselle}
and references therein.) This spectrum is obtained by the canonical light
cone quantisation of a free bosonic string in $D=26$
\cite{GGRT_Arvis}. 
Using the formula in other dimensions is known to be inconsistent: while world-sheet 
Lorentz invariance is respected, bulk Lorentz invariance is not (except, 
interestingly enough, in $D=3$
\cite{townsend})
but is restored for long enough strings. The fact that it does in practice
provide a very good approximation to the numerically determined spectrum
\cite{AABBMT11,AAMT13,AABBMT},
for reasons that were originally mysterious but are now much better understood
\cite{OAZK,OA,SD_long,SD_short},
makes it our first choice of comparison.

The spectrum is as follows. Upon quantisation the transverse fluctuations
of the string become massless `phonons' that live on the background
closed string. In $D=2+1$ there is only one transverse direction so the phonon
is a single scalar field. Since the string is periodic with length $l$,
the phonons would have momenta  $|p|=2\pi k/l$, with $k$ an integer, if
they were free. (Which we shall assume, albeit wrongly, for the moment.) 
The energy levels are then simply related to the number of 
phonons as follows. Label the positive momentum phonons as left-movers ($L$) 
and the negative ones as right-movers ($R$). Let $n_{L(R)}(k)$ be the number 
of left(right) moving phonons of momentum $|p|=2\pi k/l$. Let the total 
energy of the left(right) moving phonons be $2\pi N_{L(R)}/l$, then
\begin{equation}
N_L = \sum_k  n_L(k)k, \qquad
N_R = \sum_k  n_R(k)k.
\label{eqn_NLR}
\end{equation}
Let $p=2\pi q/l$ be the total longitudinal momentum of the string, 
then since it is the phonons that provide the momentum we have
\begin{equation}
N_L - N_R = q.
\label{eqn_NLRmom}
\end{equation}
Then in terms of these quantities the expression for the energy levels of the 
Nambu-Goto string in $D=2+1$ is
\cite{GGRT_Arvis}
\begin{equation}
E^2_{N_L N_R,NG}(q,l)
=
(\sigma l)^2 
+
8\pi\sigma \left(\frac{N_L+N_R}{2}-\frac{1}{24}\right)
+
\left(\frac{2\pi q}{l}\right)^2 ,
\label{eqn_EnNG}
\end{equation}
where the $1/24$ term can be though of as arising from the oscillator
zero-point energies, and $\sigma$ is the string tension. 
These energy levels have, in general, a degeneracy
which depends on the number of ways the  particular values of $N_L$
and $N_R$ can be formed from the $n_L$ and $n_R$ in eqn(\ref{eqn_NLR}).

At this point we note that the energy in eqn(\ref{eqn_EnNG}) is not
the simple sum of phonon energies plus the energy of the background
flux tube. That is to say, our interpretation of the states in terms of
free phonons, with corresponding momenta, is a fiction. Nonetheless it
provides a convenient mnemonic for describing the states, and so we shall
usually maintain that language in our discussion below.

Let us take the closed string to wind around the spatial $x$-torus. Let
the local transverse displacement of the string be $h(x)$. Then under the
transverse parity $P:(x,y) \to (x,-y)$, we have $h(x) \to -h(x)$ and so 
$a_k \to -a_k$, where $a_k$ is the creation operator for a phonon of 
momentum $2\pi k/l$. It arises from taking the Fourier decomposition of 
$h(x)$ and quantising it. The states are obtained by applying 
these creation operators to the `vacuum' of the world sheet theory.
So all this implies that the (transverse) parity of a state is simply 
given by the total number of phonons:
\begin{equation}
P = (-1)^{number\ of\ phonons}.
\label{eqn_Pd3}
\end{equation}
We can also consider a longitudinal parity transformation, $(x,y) \to (-x,y)$.
Since this will also reverse the direction of the flux, hence giving an 
operator with (usually)  zero overlap onto the original operator,
we always couple it with charge conjugation to undo the flux reversal.
This gives us the combined operator $P_r$. Now under $P_r$
the individual phonon momenta are reversed, as is the overall 
momentum. Thus this quantum number is only useful in the $p=0$ sector
and here the lightest non-null pair of states with $P_r=\pm$ is
$\{a_{2}a_{-1}a_{-1} \pm a_{1}a_{1}a_{-2}\}|0\rangle$ and is quite
heavy. In practice this means that this quantum number is of minor utility 
in our calculations and we shall mostly ignore it in the labelling
of our states (but will return to it later). We do not consider transverse 
momentum since that brings nothing new. So the useful quantum numbers for
a string of length $l$ are the longitudinal momentum $p$, the (transverse) 
parity $P$ and occasionally the longitudinal parity $P_r$. 

As shown in
\cite{AABBMT11}
the spectrum in eqn(\ref{eqn_EnNG}) provides an excellent description
of the flux tube spectrum even down to small values of $l\surd\sigma$.
That flux tubes should follow this very simple string spectrum was
unexpected, but is now better understood. First of all it was shown in
\cite{OA,OAZK}
(see also the earlier work in
\cite{ML,JD})
that if one considers a general effective string action and considers
strings long enough that one can express the energy as a power series
in $1/l^2\sigma$, then the first 3 correction terms to the leading
linear piece are universal. Here the universality class includes all
theories where the only massless modes on the world sheet
are the Goldstones associated with the spontaneous breaking of
translation symmetry. In addition the coefficients of these $1/l,\,1/l^3$
and $1/l^5$ terms in the expansion of $E_n(l)$ around $\sigma l$ are
precisely the same as one obtains, to that order, by expanding
the expression for $E_{N_L,N_R}(q,l)$ in eqn(\ref{eqn_EnNG}) in powers
of $1/l$. This explains why, in general, longer flux tubes are
well described by the Nambu-Goto spectrum. Why this should continue
to be the case for much shorter flux tubes has been elucidated in
\cite{SD_short}.
One can show that the $D=1+1$ world sheet theory is, to a first approximation, 
integrable. The physics of this integrable theory is determined 
entirely by the phase shift in the elastic scattering of two phonons.
This phase shift must be chosen to reproduce the finite volume Nambu-Goto
spectrum in $D=26$ and at lower $D$ one then obtains the Nambu-Goto
spectrum in  eqn(\ref{eqn_EnNG}) if one calculates in the asymptotic 
Bethe Ansatz approximation.
One can write the effective world sheet action that reproduces this
phase shift, and one can then incorporate new higher order terms or
terms corresponding to new massive fields as needed. One can then
reverse the process to calculate the corresponding scattering phase
shifts. By working backwards and searching for the classic signal of a 
phase shift passing through the value of $\pi$, this provides a powerful 
framework for determining whether the numerically determined flux tube 
spectrum contains the contribution of massive resonant states
\cite{SD_massive}.

As remarked above, we now know from the universality arguments
\cite{OA,OAZK}
that the leading correction at large $l$ to the Nambu-Goto expression
for the energy $E_n(l)$ is at $O(1/l^7)$. If, as we expect, the dynamics
of confining flux tubes is determined by an effective string action, then
we should be able to see such a correction in our numerically determined
spectrum. To see it one needs to be at large enough $l$ that one has a 
convergent expansion of $E_n(l)$ in powers of $1/l^2$. While it is not
obvious what the lower limit for such $l$ might be, a plausible estimate
is given by the value of $l$ at which the expansion of the Nambu-Goto
expression in eqn(\ref{eqn_EnNG}) begins to diverge. For $p=0$ this 
corresponds to the restriction 
\begin{equation}
l\surd\sigma
\geq
\left|4\pi\left(N_L+N_R-\frac{1}{12}\right)\right|^{\frac{1}{2}}.
\label{eqn_NGcvgce}
\end{equation}
For the lowest few $p=0$ energy levels this corresponds to 
$l\surd\sigma \gtrsim 1.05,\, 4.91,\, 7.02, \cdots $ respectively 
and apart from
the ground state the corresponding energies for such $l$ are quite large.
Since the numerical determination of the energies becomes less precise
the larger the energy (see below) this limits our numerical determination 
of the leading non-universal power correction to the ground state -- 
at least for now. In our earlier study of the $SU(6)$ flux tube spectrum
\cite{AABBMT11}
we attempted to do this, but were only able to show that the power of the
leading correction is $\geq 5$. One purpose of this paper is to do
better than that.

One of the striking features of the fundamental flux tube spectrum
calculated in
\cite{AABBMT11}
is the absence of any sign of massive modes in the spectrum, even at
smaller $l$ where the energy gaps begin to be quite large. This is
surprising since 
one usually thinks of the string tension as being dynamically generated
by the bulk fields which have a mass gap. This motivated the numerical 
study of higher representation flux tubes such as $k$-strings
\cite{AAMT13}. 
Since the $k=2$ flux tube may be regarded as a bound state of two $k=1$ 
(fundamental) flux tubes, there must be a massive mode associated 
with the binding which should manifest itself in the flux tube spectrum. 
And indeed the energy of the first excited $k=2$ state appears to be
consistent with a massive excitation on the $k=2$ flux tube
but a naive `by eye' judgement, as in
\cite{AAMT13},
is too subjective to be at all convincing.
The recent  analysis of the same data using the Thermodynamic Bethe Ansatz 
formalism to relate the observed energy spectra to phase shifts 
\cite{SD_massive}
shows a phase shift passing through $\pi$ and hence, almost unambiguously, 
a corresponding massive mode. So a second focus of this paper is to add to
what we can infer about that massive mode.

\section{Calculating flux tube energies}
\label{section_fluxtubes} 

Our lattice calculations are standard and so we provide only a minimal
sketch here. A more complete discussion can be found in our earlier papers
\cite{AABBMT11,AAMT13}. 

\subsection{lattice setup}
\label{section_latticesetup}

Our lattice field variables are $N\times N$ $SU(N)$ matrices, $U_l$, residing 
on the links $l$ of the periodic $L^2_s L_t$ lattice, whose spacing we
label  $a$. The Euclidean path integral is 
\begin{equation}
Z=\int {\cal{D}}U \exp\{- \beta S[U]\} ,
\label{eqn_Z}
\end{equation}
where ${\cal{D}}U$ is the Haar masure and we use the standard plaquette action,
\begin{equation}
\beta S = \beta \sum_p \left\{1-\frac{1}{N} {\mathrm{ReTr}} U_p\right\}  
\quad ; \quad \beta=\frac{2N}{ag^2}.
\label{eqn_S}
\end{equation}
Here $U_p$ is the ordered product of link matrices around the plaquette $p$. 
We write  $\beta=2N/ag^2$, where $g^2$ has dimensions of mass and becomes the 
continuum coupling when $a\to 0$. 

\subsection{flux tube energies}
\label{subsection_calculations} 

To calculate the flux tube spectrum, we calculate correlation functions
of a suitable set of lattice operators $\{\phi_i\}$. Expanding the correlators 
in terms of the energy eigenstates, $H | n \rangle = E_n | n \rangle$, and 
expressing $t=an_t$ in lattice units, we have
\begin{equation}
C_{ij}(t) 
=
\langle \phi_i^\dagger(t)\phi_j(0) \rangle
=
\langle \phi_i^\dagger e^{-Han_t} \phi_j \rangle
=
\sum_k c_{ik} c^\star_{jk} e^{-aE_k n_t} ,
\label{eqn_cortomass}
\end{equation}
where $c_{ik} = \langle vac | \phi^\dagger_i | k \rangle$. We now perform a 
variational calculation on $e^{-aH}$. Suppose $\phi = \psi_0$ maximises 
$\langle \phi^\dagger(a)\phi(0) \rangle/\langle \phi^\dagger(0)\phi(0) \rangle$
over the vector space spanned by the $\{\phi_i\}$, then $\psi_0$ is our best 
estimate of the wave-functional of the ground state. Repeating this calculation 
over the basis of operators orthogonal to $\psi_0$ gives us  $\psi_1$,
our best estimate for the first excited state. And so on for the
higher excited states. We then calculate the correlators of the $\psi_i$
which, if our original basis is well chosen, have a good overlap
onto the corresponding state $ | i \rangle$ and will be dominated even
at small $t$ by that state,
$\langle \psi_i^\dagger(t)\psi_i(0) \rangle \propto \exp(-E_i t)$,
so that we can accurately estimate the desired energy $E_i$.
We can clearly do this for each set of quantum numbers separately.

The states that we are interested in are loops of flux closed around
the $x$-torus. Thus our operators will also wind around the $x$-torus.
The simplest such operator is the Polyakov loop
\begin{equation}
l_p(n_t) =  \sum_{n_y} \mathrm{Tr} 
\left\{\prod^{L_x}_{n_x=1} U_x(n_x,n_y,n_t)\right\} ,
\label{eqn_poly}
\end{equation}
where we take the product of the link matrices in the $x$-direction 
around the $x$-torus, $(x,y,t)=(an_x,an_y,an_t)$, and we sum over $n_y$ 
to produce an operator with zero transverse momentum, $p_\perp = 0$. 
(Recall that  $p_\perp \neq 0$ is not interesting.)
This operator is $x$-translation invariant and so has zero longitudinal
momentum, $p=0$. It also has positive parities, $P=P_r=+$. To access
other quantum numbers we need to add `kinks' to this simple operator.
To have a good overlap onto the light physical states we need to use
operators that are `smeared' on physical length scales, achieved
here using a simple spatial `blocking' algorithm
\cite{AABBMT11,MT_block}.
One set of such operators that we use is explicitly shown in Table 2 of
\cite{AABBMT11}.
We have also constructed and used a second set of operators that 
differs in detail but is morally the same. These sets of `kinky' and
`blocked' operators provide the basis $\{\phi_i\}$ for our
variational calculation of the flux tube spectrum.

Our $SU(N)$ matrices are in the  $N\times N$ fundamental representation 
so the trace in eqn(\ref{eqn_poly}) is in the fundamental,
and so $l_p(n_t)$ should project onto flux tubes carrying fundamental flux. 
To project onto a flux tube carrying flux in another representation we 
take the corresponding trace, using standard expressions for the latter
\cite{AAMT13,BLMT01}.

\subsection{errors}
\label{subsection_errors} 

To calculate $E_i$ from $\langle \psi_i^\dagger(t)\psi_i(0)\rangle$, we need
to go to large enough $t$ that the correlator is dominated by 
$\exp(-E_i t)$. To determine when that happens we define an effective
energy
\begin{equation}
aE_{eff}(n_t)
=
- \ln \left\{
\frac{\langle \psi_i^\dagger(n_t)\psi_i(0) \rangle}{\langle \psi_i^\dagger(n_t-1)\psi_i(0) \rangle}
\right\}.
\label{eqn_Eeff}
\end{equation}
Once a single exponential dominates the value of $\langle \psi_i^\dagger(t)\psi_i(0)\rangle$,
then $aE_{eff}(t=an_t)$ will become independent of $n_t$ and  $aE_{eff}$ will
equal the desired energy $aE_i$. So identifying such an `effective energy 
plateau' is the essential step in calculating $E_i$. 

In pure glue Monte Carlo calculations the statistical errors on the 
correlators are roughly independent of $t=an_t$ while the average value 
from which we extract the energy decreases roughly exponentially in $t$. 
Thus we need the  `effective energy plateau' to set in at small $n_t$, if 
our calculation is to be usefully accurate, and this requires that the 
operator $\psi_i$ has a very good overlap onto the true wave-functional. 
It is also clear that the larger the value of $aE_i$, the more rapidly 
the statistical error on $aE_{eff}(n_t)$ will grow with $n_t$. The corresponding 
systematic error is that we then take the plateau to start at a value 
of $n_t$ that is too small and so we over-estimate the value of $aE_i$. 
(The diagonal correlator is a sum of exponentials with positive coefficients, 
so  $aE_{eff}(n_t)$ should decrease monotonically with $n_t$.) This important 
systematic error will therefore grow  as we consider higher excited states 
at fixed $l$, or if we go to larger $l$ for a fixed eigenstate such as
the ground state.

For excited states the corresponding correlator may contain a small admixture 
of lighter states that will eventually dominate as  $n_t$ grows. In this case
what we are looking for is an effective energy plateau that exists over a
limited range of $n_t$. At some point this procedure will become ambiguous, 
and even more so if the state is unstable with a non-zero decay width. It
is therefore clear that the most reliable energy determinations will be those
of the ground states in the various quantum number sectors.

It is clear that at the foundation of our calculation must be a reliable 
identification of the  effective energy plateaux. We have shown examples in
\cite{AABBMT11}
for fundamental flux tubes, and in
\cite{AAMT13}
for flux in higher representations, so in this paper we will limit ourselves 
to displaying only a few examples where one has good reason to worry whether 
a flux tube actually exists, such as the adjoint flux tube.

\section{Leading non-universal correction}
\label{section_nonuniversal}

As remarked above, if we expand the energy levels, $E_n(l)$, of
$D=2+1$ closed flux-tubes in powers of $1/l^2\sigma$, then 
the first non-universal term is expected to appear at
$O(1/l^{\gamma\geq 7})$ relative to the leading linear term, $\sigma l$
\cite{OAZK,OA}
(see also
\cite{SD_long}).
Since the universal terms are the same as the terms one obtains
by expanding the Nambu-Goto spectrum in eqn(\ref{eqn_EnNG}), we
can write the predicted large-$l$ spectrum as
\begin{equation}
E_n(l,p) = E_{n,NG}(l,p) + \frac{c}{(l\surd\sigma)^{\gamma\geq 7}} + ...
\label{eqn_Enlargel}
\end{equation}
where the power $\gamma$ is an odd integer, and $c$ contains a factor
of $\surd\sigma$ and so has dimensions of energy.  
Since the expansion of $E_{n,NG}(l)$ in powers of $1/l$ only converges (if 
we ignore the momenta) for $l^2\sigma > l_0^2\sigma \sim 4\pi (N_L+N_R-1/12)$,
it is only for lengths $l \gg l_0$ that
the prediction in eqn(\ref{eqn_Enlargel}) is expected to be valid.
These are large values of $l$ if $N_L+N_R \neq 0$ and our current
numerical calculations do not have the statistical accuracy needed
to identify such a correction at such large $l$. It is only for the
ground state, $N_L=N_R=0$, that we can hope to identify such a correction.
Indeed, in this case the expansion of $E_{NG}(l)$ is valid for
$l_0^2\sigma > \pi/3$ which includes the full range within which
the theory is confining for $N\geq 3$. In
\cite{AABBMT11}
we analysed the ground state energy for $SU(6)$ at a lattice spacing
$a\surd\sigma \simeq 0.086$, where any lattice spacing corrections
will be very small, and concluded that a power correction as in
eqn(\ref{eqn_Enlargel}) must have $\gamma\geq 5$ but our statistical
accuracy did not enable us to unambiguously confirm the theoretical
prediction that  $\gamma\geq 7$. 
One of our main goals in this paper is to improve upon that
earlier result. This has involved making additional calculations in $SU(6)$ 
at the same value of $\beta$, and new calculations in $SU(8)$ and in $SU(4)$ 
at approximately the same value of $a\surd\sigma$.

We begin with our calculation of the spectrum in $SU(8)$ at $\beta=306.25$. 
The string tension is essentially determined by the ground state energy,
$E_0(l)$, at larger $l$ while the non-universal correction term will only
be visible (given our errors) at very small $l$. Once we have determined
$a^2\sigma$ we can calculate $E_{0,NG}(l)$ and plot the difference
$E_{0}(l)-E_{0,NG}(l)$ thus exposing, in Fig.\ref{fig_DENGpow9_su8},
deviations from Nambu-Goto at small $l$. If we assume that the correction
is an odd inverse power of $l$, as in eqn(\ref{eqn_Enlargel}), then we find
that the best fit with a power is $\gamma=9$, as displayed in the figure.
In addition while $\gamma =7$ also gives an acceptable fit, we can exclude
any value $\gamma \leq 5$, thus confirming the theoretical prediction. 

Since this is our largest value of $N$, and so our `best' calculation
in terms of minimising various systematic uncertainties, we will go into
the statistical analysis in more detail. As is apparent from 
Fig.\ref{fig_DENGpow9_su8} only the lowest 4 or 5 values of $l$
are relevant for the $1/l^\gamma$ fit. To avoid the $\chi^2$ contribution
of these points being swamped by the higher $l$ points, we perform
a best fit at each $\gamma$ to all the data, but then only consider the
$\chi^2$ contribution of these lowest 4 (or 5) points. Since $a^2\sigma$
is effectively determined by the larger $l$ points, we are effectively
determining 2 parameters, $c$ and $\gamma$, using this `reduced' $\chi^2$
and so we have 2 (or 3) degrees of freedom. From this we extract the
statistical p-values shown in Figs.\ref{fig_pvalue1-4_su8},\ref{fig_pvalue1-5_su8}.
The solid points come from using the energies listed in Table~\ref{table_su8Q0},
where we have used time-correlated fits to the correlation functions for $l\leq 20$
and non-correlated fits for $l>20$. 
We also show the result of using correlated fits at all $l$ (open circles)
or non-correlated fits at all $l$ (open boxes). The conclusion, $\gamma \geq 7$, 
is the same with all these choices, irrespective of whether we use the lowest 
4 or 5 values of $l$ in the analysis.

Our second calculation involves extending the SU(6) calculation at $\beta=171$
that was reported in
\cite{AABBMT11}.
The ground state energies are listed in Table~\ref{table_su6Q0}. As in $SU(8)$
the values for $l\leq 20$ have been obtained from time-correlated fits to the
correlation functions, while the $l>20$ fits are time-uncorrelated. Again as in
$SU(8)$, we fit all values of $l\geq 14$ but consider only the $\chi^2$ of the
lowest 4 or 5 points in obtaining the p-values displayed in Fig~\ref{fig_pvalue_su6}.
We see that while this anaysis is consistent with $\gamma\geq 7$ it certainly does
not exclude $\gamma=5$ and so is unable to provide an unambiguous test
of the theoretical predictions. This motivated the $SU(8)$ calculation we have
summarised above. The essential difference between our $SU(6)$ and $SU(8)$
calculations is that in $SU(8)$ $l=14a$ turns out to be slightly closer 
to the deconfining transition than in $SU(6)$, so the deviation is larger and
provides a more severe constraint on the power fit. At the same time $N$ is larger
so the strong first order `deconfining' transition is more abrupt, its presence
has less effect when one is close to it, and the world-sheet dynamics is less
likely to be compromised when very close to the phase transition. 

We turn now to the ground state energies in $SU(4)$ at $\beta=74$.  
These are listed in Table~\ref{table_su4Q0}. We extract the deviation from
Nambu-Goto just as in $SU(8)$ and plot the result in Fig.\ref{fig_DENGpow7_su4},
where we show a fit with an $O(1/l^7)$ correction. The  p-values corresponding 
to fits with various values of $\gamma$ are shown in Fig.\ref{fig_pvalue_su4},
and we see that the evidence for $\gamma \geq 7$ is strong. In this
$SU(4)$ calculation (unlike the $SU(8)$ one) we have also calculated the
higher representation $k=2$ ground state which we plot in Fig.\ref{fig_DENG_gsk2A_su4}.
Although the errors are larger, the deviations from Nambu-Goto are much larger,
and the power fit is very strongly constrained, as we see in Fig.\ref{fig_pvalue_su4k2A}.
Both the fits to the lowest four values of $l$, and the fits to all $l$, 
completely constrain the value of the inverse power to $\gamma=7$. Since
this is such a strong result and since we know that in $SU(4)$ the `deconfining'
transition is only weakly first order
\cite{TcSU4}
so that the nearby $l/a=13$ energy might be influenced by it, 
it is worth asking what happens if drop this $l/a=13$ value from the fit. Doing 
so we obtain a set of $p$-values also shown in  Fig.\ref{fig_pvalue_su4k2A}.
While the fit to the power $\gamma$ is now less constrained, so that  $\gamma=7$
and 9 are both allowed, a value of $\gamma<7$ is still completely excluded.

We conclude from these calculations that the leading correction to
Nambu-Goto can be fitted by a power as in eqn(\ref{eqn_Enlargel})
and that the power is constrained to be 
\begin{equation}
\gamma \geq 7  ,
\label{eqn_gammaN}
\end{equation}
as expected from the analysis of the leading non-universal contribution in
\cite{OAZK,OA}.

All the above has assumed that the deviation from Nambu-Goto at very
small $l\surd\sigma$ is given by an (effective) power correction $O(1/l^\gamma$).
Since the deviation is only significant (given the errors) for the 4 or 5
smallest values of $l$, it is obvious that we have to make an assumption
about the functional form of the correction to obtain any useful constraint.
Having said that we will briefly explore one plausible alternative to a power-like
correction, which is an exponential, $O(e^{-ml})$. One could expect such a
correction to $\sigma$ to arise from the emission and absorption of the lightest
particle in the bulk gauge theory, in which case $m$ should be the mass gap.
And in that case the coupling should decrease with $N$, so that the coefficient of
the exponential correction $\to 0$ as $N\to \infty$. More speculatively, an 
exponential might also arise as the correction to the expansion of $E(l)$ in powers
of $1/l^2\sigma$ if that expansion were asymptotic.

It turns out that fitting a correction of the form  $\propto e^{-ml}$ works
remarkably well, as we see in Figs.\ref{fig_DENGk2Aexp_su4},\ref{fig_DENGexp_su8}. 
In fact, it is easy to see, as we now show, that this follows automatically if a 
powerlike fit works well and that power is large. If the power is large then 
the correction decreases rapidly as $l\surd\sigma$ increases. With finite errors, 
and given a lower bound provided by the `deconfining' transition, the range of $l$ 
where the deviation from Nambu-Goto is visible will be small, as we see in
Figs.\ref{fig_DENGpow9_su8},\ref{fig_DENGpow7_su6},\ref{fig_DENGpow7_su4}.
Suppose that the relevant range is $l\in [l_1 - \delta_1,l_1 + \delta_1]$ with 
$\delta_1 \ll l_1$ as will be the case if the range is narrow. Then we can write 
the power correction at $l= l_1 + \delta l \in [l_1 - \delta_1,l_1 + \delta_1]$ as 
\begin{eqnarray}
\frac{1}{l^\gamma} 
&=& 
\frac{1}{(l_1 + \delta l)^\gamma} = e^{-\gamma \ln (l_1 + \delta l)}\nonumber \\
&=&
c  e^{-\gamma \ln (1 + \frac{\delta l}{l_1})} \nonumber \\
&= & 
c^\prime  e^{-\frac{\gamma}{l_1} \delta l} \left(1 + 
\frac{\gamma}{2}\left(\frac{ \delta l}{l_1}\right)^2 + ...\right)\nonumber \\
&\simeq&
c^{\prime\prime} e^{-\frac{\gamma}{l_1} l}
\label{eqn_powtoexp}
\end{eqnarray}
for our typical $ \delta l/l_1$ and $\gamma$, and within finite errors. 
Thus a good fit with a $\propto 1/l^\gamma$ correction necessarily implies a good 
fit with a  $\propto e^{-ml}$ correction, where $m\simeq \gamma/l_1$. Indeed
the masses in our exponential fits do satisfy this expectation. 

Of course the above argument works both ways; if an exponential correction
is the correct dynamics then this can imply that a powerlike correction
will also fit well. However we note that while the best value of the power
is $\gamma = 7,9$, which is in line with the theoretical expectations, the values 
of $m$ obtained using an  $e^{-ml}$ correction do not coincide with the
bulk mass gap, and the coefficient, although it decreases by about $35\%$ 
between $SU(4)$ and $SU(8)$, does not decrease anything like $1/N^2$.
It therefore appears quite implausible that the correction is really
exponential rather than powerlike. 

\section{Massive modes and $k$-strings}
\label{section_massive} 

A striking feature of the spectrum of fundamental representation confining
flux tubes in $D=2+1$ $SU(N)$ gauge theories is the lack of any sign of
massive modes
\cite{AABBMT11}.
This is in contrast to $D=3+1$ 
\cite{AABBMT10}.
This is puzzling if one believes that the confining flux tube
has some non-zero but finite `intrinsic' width, since this would
imply the existence of corresponding massive modes. Motivated by
this puzzle we calculated in 
\cite{AAMT13}
the spectrum of flux tubes carrying flux in higher representations
than the fundamental. Any such flux tube, if stable, can be thought
of as a bound state of some number of (anti)fundamental flux tubes, and
the massive modes inducing that binding should be encoded in the
flux tube spectrum in some way. Particularly relevant are $k$-strings
which one can think of as bound states of $k$ fundamental flux tubes.
Two flux tubes carrying flux $k,k^\prime \leq N/2$ can be shown not
to  mix if $k\neq k^\prime$ using a standard centre symmetry argument.
From earlier lattice calculations
\cite{BLMT01,sigma_k2}
we know that the lightest $k$ string has a string tension 
$\sigma_k < k\sigma_f$ for $k>1$ so it will be stable against decay to
$k$ fundamental flux tubes. In particular the ground state $k=2$ and $k=3$ 
flux tubes are absolutely stable. Since these are stable bound states,
the finite binding energy per unit length must be associated with some
massive mode and it is  interesting to see if one can identify it within
the spectrum. In 
\cite{AAMT13}
we found that the energy of the first excited $k$-string state had a 
behaviour with $l$ consistent with that of a massive mode, but that this
interpretation was  far from unambiguous given the heuristic nature of our
analysis. This uncertainty has subsequently been removed by the approach in
\cite{SD_massive}
that uses a world-sheet effective field theory framework that
allows one to work back from the spectrum to the scattering phase
shifts of the phonons living on the world sheet. These calculations
\cite{SD_massive}
have confirmed  the existence of a massive resonance on the world sheet
of $k=2$ and $k=3$ flux tubes, and that its mass, $M_k$, is indeed very 
close to that of the nearly constant gap between the ground and first 
excited flux tube states. 

The above massive mode calculations used the $SU(6)$ $k$-string spectra in
\cite{AAMT13}.
It is of interest to ask if this mode only knows about the world-sheet
theory, and so knows only about $\sigma_k$, or if it knows about the
bulk gauge theory. In the former case $M_k/\surd\sigma_k$ should be some
fixed number independent of which $SU(N)$ group generates the bulk physics.
Since $\sigma_k/\sigma_f$ varies with $N$ this predicts that $M_k/\surd\sigma_f$
should vary with $N$. In the latter case it should be $M_k/\surd\sigma_f$
or  $M_k/m_G$, where $m_G$ is the bulk theory mass gap, that are weakly
dependent on $N$ (up to $O(1/N^2)$ corrections). To address
this question we will compare what one finds in $SU(6)$ and $SU(4)$, on the
lattices and at the $\beta$-values discussed in Section~\ref{section_nonuniversal}.

To begin with, we plot in Fig.\ref{fig_Ek2A_q0P+_su6} the lightest few states
with positive parities and $p=0$, in the spectrum of $k=2A$ flux tubes 
in $SU(6)$. We see that for $2.5 \lesssim l\surd\sigma\lesssim 5.5$  the
gap between the ground and first excited states is roughly constant and so we 
interpret the first excited state as a massive mode on the ground state flux tube. 
The next excited state is quite close to the first excited Nambu-Goto level
and approaches it as we increase $l$, so we would interpret that as being
(predominantly)  the Nambu-Goto state with one left and one right moving massless 
phonon on the flux tube. There appears to be a level crossing at 
$l\surd\sigma\sim 5.5$ (with level repulsion in the neighbourhood) so by the time 
we are at  $l\surd\sigma\sim 6$ we would interpret the first excited state as
being the Nambu-Goto state with one left and one right moving massless phonon,
while the second excited state is now the massive mode on the flux tube. 
Some of the complication of the spectrum may well arise from a state 
that contains two massive modes. 
We also show, in Fig.\ref{fig_Ek2A_q0P+_su4}, a corresponding plot for $SU(4)$.
The level crossing appears to occur a little earlier, at
$l\surd\sigma \sim 5$, but otherwise it is similar to the $SU(6)$ spectrum.
Finally we display in Fig.\ref{fig_Ek3A_q0P+_su6} the spectrum of the
lightest $k=3A$ flux tube states in $SU(6)$. Here the features identified in the 
$k=2A$ spectrum are even more clearly on display: a near-constant gap to the first 
excited state, a separate stringy state approaching the first excited
Nambu-Goto level, together with level crossing at larger $l$ and associated
level repulsion. As we see from Table~\ref{table_suNsigmaTc} the binding of the
$SU(6)$ $k=3$ ground state, as measured by the ratio $\sigma_{k=3}/3\sigma_f \simeq 0.61$, 
is stronger than that of the $SU(4)$ $k=2$ ground state, where
$\sigma_{k=2}/2\sigma_f {\simeq} 0.68$, or of the $SU(6)$ $k=2$ 
ground state, where $\sigma_{k=2}/2\sigma_f {\simeq} 0.81$, and
this may be a reason for the particular clarity of the massive mode signal we
see in  Fig.\ref{fig_Ek3A_q0P+_su6}.

The energy gap between ground and first excited $k$-string states can be extracted
from the energies in Table~\ref{table_su4su6Ek2A3A}. We plot the $k=2A$  energy gap 
in units of the world-sheet string tension, $\sigma_{k=2A}$, versus the length $l$ 
in Fig.\ref{fig_DE1E0k2A_sigk2A_n4-16}. In addition to the calculations listed
in Table~\ref{table_su4su6Ek2A3A} we have also included some values obtained
in $SU(8)$, $SU(12)$ and $SU(16)$ as part of our glueball spectrum calculation in
\cite{AAMT16}.
The $SU(8)$ and $SU(12)$ calculations are at values of $a\surd\sigma_f$ similar
to those of our $SU(4)$ and $SU(6)$ calculations, while the $SU(16)$ lattice
spacing is somewhat larger (but we do not expect this to matter). We contrast this
with the plot in Fig.\ref{fig_DE1E0k2A_m0pp_n4-16} where the same energy is
expressed in units of the bulk theory (infinite volume) mass gap. We observe that
the variation with $N$ in the latter plot is much smaller than that in the former.
This suggests, quite strongly, that the mass of the massive mode is determined
by the physics of the bulk $SU(N)$ theory, which is indeed what one might expect
if it reflects the binding of two fundamental strings into a $k=2$ string.

In  Fig.\ref{fig_DE1E0k2Ak3A_m0pp_n6} we compare the energy gaps of $k=2$ and
$k=3$ flux tubes in our $SU(6)$ calculation. We see that they indicate
an identical massive mode when expressed in units of the bulk mass gap, 
which again would not be quite the case if we used the corresponding 
$\surd\sigma_k$ to scale the energy gaps. (But not a strong effect because the
$k=2$ and $k=3$ string tensions happen to be quite close here.)

\section{Orthogonality of representations}
\label{section_overlaps} 

If we have a source in a (non-singlet) representation $\cal{R}$ of the $SU(N)$ group,
it will be linearly confined so that the flux between it and a corresponding distant
anti-source will be carried by a flux tube in the same representation  $\cal{R}$, and
with a string tension $\sigma_{\cal{R}}$. Or at least this is what would happen if there
were no screening. In reality, however, the vacuum contains a condensate of adjoint
gluons which will presumably screen such a source down to the representation
$\cal{R}^\prime$ which has the smallest associated string tension. Since adjoint
screening leaves a source in the same $k$-string sector
\cite{AAMT13}, 
each such sector (usually referred to as a sector of fixed $\cal{N}$-ality, with 
${\cal{N}}=k$) contains its own absolutely stable minimum energy flux tube, which 
is what we are usually referring to when we speak of a `$k$-string'. Since one believes
the gluon condensate to be strongly coupled, one would naively expect the screening
to have a large probability as soon as it is energetically favoured, so that the
$\cal{N}$-ality of the flux would be its only useful label. For a flux tube
which is not tied to sources but which is closed around a spatial torus of length $l$,
such a screening will be energetically favoured for all $l>l_c=1/T_c$ (possibly
qualified at small $l$ by corrections to the linear piece of the energy). So
what we would expect to obtain from the $t$-dependence of a correlator of operators
representing closed strings in any representation $\cal{R}$ of a $k$-sector is
simply the energy of the associated $k$-string. And if we consider two operators in
different representations $\cal{R}$ and $\cal{R}^\prime$ but in the same $k$-sector,
then we would expect their overlap to be substantial.

In practice one finds that these naive expectations are largely contradicted
by actual calculations. In effect the screening appears to be a very weak process.
For example, if one looks at closed flux tubes in the $k=2A$ and $k=2S$ representations,
which are the symmetric and antisymmetric pieces of $f \otimes f$, one finds a reasonably
well defined $k=2S$ flux tube energy which is much larger than the corresponding  $k=2A$
energy. Compare, for example,  Fig.1 and Fig.11 in
\cite{AAMT13}.
Similarly in the $k=3$ sector: compare Fig.15 and Fig.19 in
\cite{AAMT13}.
Moreover one finds that typical overlaps between closed flux tube operators in different
representations (within the same $k$-sector) are very small, perhaps even consistent with
zero up to finite volume effects. This was already noted in, for example,
\cite{AABBMT}
(see Tables 2,3 therein) and
\cite{AAMT13}
(see Section 4.1 therein). These observations were made using individual operators
chosen from within a small basis. (This basis was designed to have a good projection onto
ground states but not onto excited states.) Here we will improve the analysis by
using our large basis of flux tube operators, and calculating the overlaps using 
the orthonormal operators
produced by our variational calculation, which are our best estimates for the actual energy
eigenstates. Moreover we will be able to compare in some detail what happens in $SU(4)$ 
and $SU(6)$ for $k=2$, and we will include  $SU(2)$ when we consider $k=0$ (i.e. the 
adjoint flux tube).

We begin with our various winding operators $l_{p,j}$ which are constructed using 
link matrices in the fundamental representation. Typically we have 17 operators at
each blocking level, and we have all blocking levels $n_b$ such that $2^{n_b -1} \leq l/a$.  
(For small $l/a$ we sometimes include a larger `transverse' blocking level.) 
Taking the trace we obtain ${\mathrm Tr}_f{l_{p,j}}$ as our basis for flux tubes in the fundamental. 
To obtain a basis in the $k=2$ symmetric, antisymmetric (2A,2S) and $k=0$ adjoint 
representations we  simply take
\begin{eqnarray}
{\mathrm Tr}_{2A}{l_{j}} 
& = &
\frac{1}{2}\left( {\mathrm Tr}_f\{l^2_{j}\} - \{{\mathrm Tr}_f l_{j}\}^2 \right) , \nonumber \\
{\mathrm Tr}_{2S}{l_{j}} 
& = &
\frac{1}{2}\left( {\mathrm Tr}_f\{l^2_{j}\} + \{{\mathrm Tr}_f l_{j}\}^2 \right) , \nonumber \\
{\mathrm Tr}_{adj}{l_{j}} 
& = &
\{{\mathrm Tr}_f l_{j}\}^2 - 1 ,
\label{eqn_kfromf}
\end{eqnarray}
where we drop the subscript $p$ on $l_p$ from now on to declutter the notation.
We then take spatial sums of these to project onto $p=0$ and positive parities. 
We can equally well consider $p \neq 0$ and negative parity.  We label these
normalised operators $\phi_{{\cal{R}},j}(t)$, suppressing other quantum number labels. 
In our earlier work
\cite{AAMT13,AABBMT}
we considered a few individual overlaps  $\phi_{2A,j}(0) \phi_{2S,j}(0)$ and showed 
they were small and rapidly decreasing as $l$ increases. However since 
$\phi_{{\cal{R}},j}(t)$ and  $\phi_{{\cal{R}},l}(t)$ are not orthonormal (in fact the
overlap for $j\neq l$ is often large) we did not make a statement about the total 
overlap between the bases. So here we shall produce an orthonormal set of
operators from the $\phi_{{\cal{R}},j}(t)$. A particularly useful choice, which 
is the one that we shall use below, is
provided by our variational procedure which produces orthonormal linear
combinations  $\psi_{{\cal{R}},j}(t)$ which are our best variational estimates
for the wave-functionals of the energy eigenstates. We shall illustrate 
our results for what happens in general with two particular cases:
the overlap between flux tubes in the $k=2$ symmetric ($2S$) and antisymmetric 
($2A$) representations, and the overlap between the adjoint flux tube and the 
(singlet) vacuum.

\subsection{$k=2A $ and $k=2S$} 
\label{subsection_k2overlap}

Consider the square of the overlap of our best (variational) operator for the  
$2A$ ground  state flux tube onto our whole $2S$ basis
\begin{equation}
{\cal{O}}_{2Ags,2S}
= 
\sum_k \left| \langle \psi^\dagger_{2A,j=0}(0) \psi_{2S,k}(0) \rangle \right|^2 ,
\label{eqn_O2Ags2S}
\end{equation}
or of our $2S$ ground state on to our $2A$ basis
\begin{equation}
{\cal{O}}_{2Sgs,2S}
= 
\sum_k \left| \langle \psi^\dagger_{2S,j=0}(0) \psi_{2A,k}(0) \rangle \right|^2 ,
\label{eqn_O2Sgs2A}
\end{equation}
or of our whole $2A$ basis on our whole $2S$ basis
\begin{equation}
{\cal{O}}_{2A,2S}
= 
\sum_k\sum_j \left| \langle \psi^\dagger_{2A,j}(0) \psi_{2S,k}(0) \rangle \right|^2 .
\label{eqn_O2A2S}
\end{equation}
Note that because the operators all have unit normalisation, these overlaps
have different maximum values, i.e. while ${\cal{O}}_{2Ags,2S} \leq 1$ and  
${\cal{O}}_{2Sgs,2A} \leq 1$, we have 
${\cal{O}}_{2A,2S} \leq {number \, of\, operators}$. Note also 
that these expressions only provide an approximation to the true overlaps 
since our basis of operators is not complete. However we believe that this
approximation is a good one since we find that highly excited states contribute
little to either overlap and our basis of operators is large enough to have a 
good overlap onto the lighter states. There are of course many other instructive
overlaps that one can consider, but for brevity we shall here focus on those 
defined in eqns(\ref{eqn_O2Ags2S},\ref{eqn_O2Sgs2A},\ref{eqn_O2A2S}).

We begin by showing in Fig.\ref{fig_k2Atotk2S_n4} the total overlap,
${\cal{O}}_{2A,2S}$, between our $2A$ and $2S$ bases as a function
of the flux tube length $l$, obtained in $SU(4)$ at $\beta=74$.
We do so for 3 bases of operators which are distinguished by the
maximum `blocking level' of the operators in the basis. Since a
blocked link at level 6 has length 32 lattice spacings, we only
include it for $l/a\geq 32$. (The reader will  notice that we have 
included blocking level 5 for $l/a\leq 16$ in  Fig.\ref{fig_k2Atotk2S_n4}.
This is not a normal blocked link, but rather a blocking level 4 link that
has been smeared further in the transverse directions.) The maximum
value of ${\cal{O}}_{2A,2S}$ is $17\times (\mathrm{number\, blocking\, levels})$,
and this is the normalisation to bear in mind when reading this plot.

We observe that at very small $l$, close to $l_c \equiv 1/T_c$, the total $2A/2S$ 
overlap is substantial. (Recall that this is the overlap squared.) However
we also see that as the length of the flux tube increases the overlap
decreases very rapidly, more or less exponentially with $l$. At the largest
values of $l$ the signal is effectively lost in the noise and our results
are consistent with  ${\cal{O}}_{2A,2S} \stackrel{l\to\infty}{\longrightarrow} 0$,
i.e. the complete absence of any screening.
An important remark is that for all our values of $l$ the basis up to blocking 
level 5 is not significantly improved by including blocking level 6, for at least 
the lightest 10 states (in the positive parity sector). By contrast, going from
maximum blocking 4 to level 5 provides a marked improvement. Thus we may consider
the basis with maximum blocking level 5 as providing a good representation of the
total overlap in a large volume.

The results with a maximum blocking level of 6 illustrate a general feature of our 
overlaps in various representations. When the blocking level is large enough that
the operator effectively wraps around the spatial volume in all directions, then 
the corresponding overlaps become substantial, even though they follow the
usual pattern of rapidly decreasing as the volume is further increased. This 
substantial overlap appears to be a finite volume effect that may well be peculiar 
to our operator construction, and is something that we need to understand better.

We turn now to our best operator for the $2A$ ground state, again
in $SU(4)$ at $\beta=74$. The overlap of
this operator onto the actual ground state is $\sim 99\%$ if we use the basis up to
blocking level 5, and so provides us with an excellent approximation to the true
overlap. We see from  Fig.\ref{fig_k2Agsk2S_n4} that the overlap of this state 
onto our $2S$ basis is large for the very shortest flux tubes. (Recall that the 
maximum value of the overlap is unity in this case.) So here there appears
to be no suppression of screening. However the screening decreases rapidly - at
least exponentially - as $l$ increases and is consistent with vanishing as
$l\to\infty$. In practice, calculating within the $k=2A$ sector provides us
with an excellent approximation to the absolute ground state in the $k=2$ sector.

Finally we show in  Fig.\ref{fig_k2Sgsk2A_n4} the same plot, but for the 
overlap of the $2S$ ground state onto the $2A$ basis. In this case the overlap 
of our best variational operator onto the ground state is only $\sim 90\%$.
This ground state has a large energy and should be unstable into a pair of 
fundamental flux tubes with equal and opposite momentum. Since 
$\sigma_{2S} \simeq 2.6 \sigma_f$ there is plenty of phase space for such 
a decay. (Moreover this process is not large-$N$ suppressed, as we see 
from eqn(\ref{eqn_kfromf}).) It is therefore worth making a brief detour 
to see if  there is in fact a $k=2S$ ground state that is sufficiently 
stable to be well defined. So we plot in Fig.\ref{fig_Eeffk2Sgs_n4} the
effective energies obtained from our best variational correlators
\begin{equation}
aE_{eff}(n_t) = -\ln\frac{C(n_t)}{C(n_t-1)} \, ,
\label{eqn_Eeffb}
\end{equation}
for various $l$ in our $SU(4)$ calculation. A well defined state
should produce a plateau in $aE_{eff}(n_t)$ starting from some
value of $t=n_t a$, indicating that the correlator is then given
by a single exponential. A state with a small decay width will produce
an apparent plateau in the values of $aE_{eff}(n_t)$ but will, for large 
enough $n_t$, decrease towards an asymtotic value given by the lightest decay 
state. If the overlap of our best operator onto the corresponding eigenstate
is not very large, the start of the plateau is pushed out 
to larger $n_t$, and will be submerged in statistical errors,
especially for the larger energies that one obtains at large $l$.
So we see in Fig.\ref{fig_Eeffk2Sgs_n4} reasonable evidence for a
plateau for the smaller values of $l$, but at larger $l$ the evidence for
a plateau is much less convincing. Since $\sigma_{2S}\to 2\sigma_f$ and
$\sigma_{2A}\to 2\sigma_f$ as $N\to\infty$ the phase space for decay 
diminishes for larger $N$, and so one might expect the $2S$ flux tube to 
become more stable. We show in Fig.\ref{fig_Eeffk2Sgs_n8} the results
of some calculations in $SU(8)$ at $\beta=306.25$ which indeed show much
clearer plateaux in $aE_{eff}$ at larger $l$, aided by the visibly better 
overlaps. (For the corresponding plot in $SU(6)$ at $\beta=171$ see Fig.11 in
\cite{AAMT13}.) 
We conclude from this detour that it is meaningful to talk of a (presumably
`resonant') $2S$ ground state and to discuss its overlap on to the $2A$ basis. 
Returning to Fig.\ref{fig_k2Sgsk2A_n4}, what we
see is very similar to what we saw for the $2A$ ground state, except that
at intermediate values of $l\surd\sigma$ the overlaps in this case are somewhat 
larger. But again what we see is consistent with rapidly  vanishing overlaps as
$l\to\infty$.

\subsection{adjoint and singlet}
\label{subsection_adjoverlap} 

As our second example we consider adjoint ($k=0$) flux tubes and their overlap 
onto the colour singlet ($k=0$) vacuum.  We work in $SU(2)$ and $SU(4)$ at
couplings $\beta=16,\, 74$ respectively, where the values of the lattice
spacing, $a\surd\sigma_f$, are roughly the same as in $SU(6)$ at $\beta=171$.
Such a flux tube can decay, for example to the vacuum or a glueball via gluon 
screening, or into a pair of fundamental and antifundamental flux tubes. Gluon 
screening is large-$N$ suppressed, but the decay into a pair of fundamental 
flux tubes is not, as one can see from eqn(\ref{eqn_kfromf}). In addition the
adjoint string tension (in units of $\sigma_f$) increases as $N$ decreases, 
so that the phase space for these decays increases at lower $N$. All this suggests 
that even if adjoint flux tubes are well-defined at larger $N$, as shown for 
example in our study of $SU(6)$ in
\cite{AAMT13},
they may not exist as meaningful states in $SU(2)$, and perhaps even in $SU(4)$.

As a first step in addressing this question, we show in 
Fig.\ref{fig_Eeffk0gs_n2} the effective masses obtained from our 
variationally determined ground state correlator in $SU(2)$. (We use vacuum 
subtracted operators, although this is usually unnecessary.) For all except 
the largest values of $l$ there is a reasonably clear plateau, at least for 
some intermediate range of $n_t$. However the overlaps of the ground states onto 
our best operators are not very good, as shown by the marked increase of 
$aE_{eff}(n_t)$ as $n_t\to 0$. Thus for the largest values of $l$ it is hard to 
infer a plateau: the signal drops into the statistical noise before one 
has clear evidence either way. Nonetheless the fact that we do see plausible 
evidence for a quasi-stable state for almost all our values of $l$ is 
striking given our earlier comments. In addition, the fact that the energy
increases with $l$ tells us that we are not dealing with a particle-like state, 
but rather with some kind of string-like state. 
Indeed the effective plateau mass for say $l=38$, where we see a clear 
plateau, is $\sim 50\%$ higher than the lightest glueball.
(See below for a more detailed analysis.) 

We show the corresponding plot for $SU(4)$ in   Fig.\ref{fig_Eeffk0gs_n4}. Here 
the overlaps are evidently very good, and the existence of a well-defined 
`flux tube' state seems evident for all our values of $l$.

We now turn to a quantitative measure of the overlap between adjoint and singlet.
The first quantity we calculate is the total overlap squared onto the vacuum  of 
our basis of adjoint operators, 
\begin{equation}
{\cal{O}}^{tot}_{adj}
= 
\sum_k \left( \langle \psi_{adj,k} \rangle \right)^2 ,
\label{eqn_adjallvac}
\end{equation}
where $\{\psi_{adj,k}\}$ are the orthonormal operators obtained from our variational 
procedure applied to our original non-orthonormal basis of adjoint operators which, 
in turn, are obtained from the fundamental ones using eqn(\ref{eqn_kfromf}). In 
Fig.\ref{fig_adjtotvac_n2} and Fig.\ref{fig_adjtotvac_n4} we show the values
of this total overlap for $SU(2)$ and $SU(4)$ respectively. As in the earlier
discussion of $k=2$ overlaps, we show separately what one obtains using a basis that
contains operators up to blocking level 5 and up to 6 (at the larger values of $l$ 
where it has been calculated). As in the $k=2$ case we observe a rapid decrease
of the overlap  ${\cal{O}}^{tot}_{adj}$ with increasing $l$, consistent with the
overlap vanishing as $l\to\infty$. At small $l$ the overlap onto the vacuum is large.
(Recall that  ${\cal{O}}^{tot}_{adj} \leq 1$.)
And again the operators with the highest blocking levels that wrap around the whole
spatial volume have a large overlap, although at any fixed blocking level, however 
large, the overlap decreases at least exponentially fast with $l$. 

A related question is what is the overlap of the ground state adjoint flux tube 
onto the vacuum. One way to approach this question is as follows. Consider the 
orthonormal vacuum subtracted operators that arise from the variational procedure,
$\psi_{adj,i} \equiv {\tilde\psi}_{adj,i}-<{\tilde\psi}_{adj,i}>$.
We take the ground state operator ${\psi}_{adj,i=0}$, which is our 
best estimate for the adjoint flux tube ground state, and plot the value
of the vacuum expection value that was subtracted from that operator 
normalised by the vacuum subtracted correlator (which we show explicitly
although its value is unity):
\begin{equation}
{\cal{O}}^{gs}_{adj}
= 
\frac{\langle {\tilde\psi}_{adj,k=0} \rangle^2}
{\langle {\psi}^2_{adj,k=0} \rangle} .
\label{eqn_adjgsvac}
\end{equation}
This quantity provides a measure of the would-be vacuum expectation value
of the adjoint flux tube. We plot this quantity in 
Figs~\ref{fig_adjgsvac_n2},~\ref{fig_adjgsvac_n4} for our $SU(2)$ and $SU(4)$ 
calculations. We see that the vacuum expectation value is very small
except at the very smallest values of $l$, and rapidly decreases as we
increase $l$. A more direct measure is provided by applying the variational 
procedure to our adjoint operators without any vacuum subtraction, and
calculating the vacuum expectation value squared of the ground state.
This in fact produces very similar results to what we see in
Figs~\ref{fig_adjgsvac_n2},~\ref{fig_adjgsvac_n4}, except that at the very
smallest values of $l$, $l/a\leq 18$, where the whole basis has a large 
overlap onto the vacuum, the variational calculation automatically picks
the vacuum as the ground state. We also note that, once again,
where the operator wraps around the whole spatial 
volume we see large overlaps. However this appears to be finite volume effect,
since at any fixed blocking level, the overlap eventually decreases rapidly 
with $l$ and indeed with $N$. So apart from this finite volume effect, we see 
that even a modestly long adjoint flux tube has a negligible probability to be 
screened to the vacuum.

In retrospect it would have been interesting to perform a similar analysis for the
overlap of the adjoint flux tube onto, say, the lightest glueball. Technically
this is trivial: it would simply involve combining in a single global basis both 
our glueball and flux tube operators. Unfortunately in the present work we did not 
calculate such an enlarged cross-correlation matrix and so are not able to 
address this question.

\section{Spectra}
\label{section_spectra}

So far we focused on the absolute ground state. Here we will discuss part of 
the larger spectrum. In particular it is interesting to see how much of the 
striking simplicity that we observed in $SU(6)$ 
\cite{AABBMT11}
appears at smaller values of $N$, such as $SU(4)$ and $SU(2)$. We also discuss 
more explicitly the continuum and large-$N$ limits.

\subsection{$SU(2)$, $SU(4)$, $SU(6)$ and $SU(8)$ with $p=0$}
\label{subsection_su2su4su6su8p0}

An amusing feature of $SU(2)$ is that the deconfining length, 
$l_c\surd\sigma = \surd\sigma/T_c$, is smaller than the length at which 
the Nambu-Goto ground state becomes tachyonic, $l_0\surd\sigma = \pi/3$. 
Earlier calculations with a small basis of operators
\cite{AABBMT11}
have indicated that in this range of $l$ the dependence of the energy
on $l$ is governed by the critical exponents of the (second-order) 
deconfining transition. In  Fig.\ref{fig_DENG_gs_su2} we show
the difference between the calculated energy and the Nambu-Goto
energy versus $l$. (Using the calculated vaue of $\sigma$ as
a convenient normalisation.) For the point with  $l<l_0$ we replace 
$E^{NG}_0 \to 0$ since its actual value is imaginary. We see that down 
to $l\surd\sigma\sim 1.5$ the calculated energy stays very close to the 
Nambu-Goto prediction, just as it does for larger $N$. However the
deviation below that is in the opposite direction to what we observe
at larger $N$. Unlike a strong first order transition, a second
order finite volume transition will affect the $l$-dependence in
the neighbourhood of the transition and this is presumably what we 
are seeing here. 

In Fig.\ref{fig_Ek1_q0Pall_su2} we show the low-lying flux tube spectrum
in $SU(2)$ and compare it to the predicted Nambu-Goto energy levels. While 
the first excited state is close to Nambu-Goto for $l\surd\sigma > 3.5$,
it is only for $l\surd\sigma \geq 5$ that the next energy level is 
more-or-less reproduced (in both degeneracy and energy) by our calculated 
spectrum. This approach to the asymptotic universal values is much slower 
than what one observes in $SU(6)$
\cite{AABBMT11}.

It is interesting that the Nambu-Goto ground state energy works so well.
It is essentially the linear piece plus pieces that can be interpreted as
the zero point energies of the Nambu-Goto excited state oscillators. 
(Or, in a different language,
\cite{SD_short},
the effect of the vacuum gas of winding phonons on the world sheet.)
It may seem puzzling that the latter appear to be closely reproduced
at small values of $l$ where the excited state energies themselves
differ greatly from the Nambu-Goto predictions. This suggests that
these differences are constrained to largely cancel each other. As one
might expect, for example, from splittings due to mixing.

In Fig.\ref{fig_Ek1_q0Pall_su4} we show the corresponding $SU(4)$ 
spectrum. The approach to Nambu-Goto with increasing $l$ is more rapid
than in $SU(2)$, but markedly less so than in the case of $SU(6)$
which is displayed in Fig.\ref{fig_Ek1_q0Pall_su6}. We also show
the $SU(8)$ spectrum in Fig.\ref{fig_Ek1_q0Pall_su8}, which is
very similar to $SU(6)$, confirming our expectation that the latter
is effectively in the large-$N$ limit.

We list the energies of the lightest four $P=+$ and the lightest two 
$P=-$ states in Tables~\ref{table_su8Q0},~\ref{table_su6Q0},~\ref{table_su4Q0} 
and \ref{table_su2Q0}.

\subsection{adjoint}
\label{subsection_adjoint}

As emphasised in Section~\ref{subsection_adjoverlap}, the adjoint flux tube 
is expected to be unstable. Nonetheless, as we have seen from the effective energy 
plots in
\cite{AAMT13}
for $SU(6)$ and in Fig.\ref{fig_Eeffk0gs_n2} for $SU(2)$ and Fig.\ref{fig_Eeffk0gs_n4}
for $SU(4)$, the ground  state adjoint flux tube appears to be well-defined in both 
$SU(4)$ and $SU(6)$ and, to some extent, in $SU(2)$. 

We saw in 
\cite{AAMT13}
that in $SU(6)$ the adjoint flux tube energy does indeed grow approximately linearly 
with $l$ at large $l$, just as one would expect for a state that is a string-like 
flux tube. In Fig.\ref{fig_Egsk0bl456b_su4} we show what happens for $SU(4)$. We plot 
separately the energy one obtains from the apparent effective energy plateau that 
one obtains with the (variationally) best ground state operator using bases
that include blocking levels up to $4,\,5,\,6$ respectively. We find that
using blocking levels up to $5$ provides a best operator with a very good 
overlap at all $l$. Including blocking level $6$ makes for an almost insignificant 
improvement in the overlap. By contrast, while using blocking levels up to $4$ 
is adequate at the very smallest values of $l$, it becomes inadequate at larger 
values, reflected in a poor plateau and a systematic upward error in the 
energy estimate. We observe that the ground state energy rapidly becomes much
larger than the lightest ($V=\infty$) glueball mass, $am_{0^{++}}\sim 0.37$, as
we increase $l$, showing that the overlap onto glueballs must be very small.
This complements our earlier observation concerning vacuum overlaps. In addition
we see that, except for the smallest values of $l$, the energy is significantly
larger than that of two fundamental loops, just as we observed for $SU(6)$ in
\cite{AAMT13}.
This strongly suggests that, once $l$ is reasonably large, what we have in $SU(4)$ 
is indeed an adjoint flux tube that has a decay width that is small enough for it 
to be well-defined.  

The analogous results for $SU(2)$ are shown in Fig.\ref{fig_Egsk0bl456b_su2}.
Once again we see an energy that increases with $l$ well beyond the mass of the 
lightest glueball ($am_{0^{++}}\sim 0.41$ in this case), which together
with our earlier results on the vacuum overlaps shows that decays which are 
large-$N$ suppressed appear to be very small in $SU(2)$ -- even though we are 
now at the smallest possible (non-Abelian) value of $N$. However, in contrast to
$SU(4)$, the energy decreases as we enlarge our basis to include ever more
smeared operators and is consistent with decreasing at any fixed $l$ to
the lightest scattering state composed of two (anti)fundamental flux tubes.
(The dashed line in  Fig.\ref{fig_Egsk0bl456b_su2}.)
That does not necessarily mean that there is no quasi-stable adjoint flux tube.
It could simply be that our highly smeared  operators effectively include good 
operators for a pair of fundamental flux loops. However the more likely 
explanation is that the decay width of any adjoint flux tube state into 
a pair of fundamental flux loops is large enough that it cannot be reasonably 
identified by our simple methods. 

In $SU(2)$ our variational calculation does not produce trial excited states 
from which we are able to extract energies by identifying an effective energy 
plateau; as we have seen, there are already difficulties with the ground state. 
In $SU(4)$, by contrast, we are usually able to estimate the energies of the
first two excited states and we plot these in Fig.\ref{fig_Ek0_q0P++_su4}
together with the ground state and also the Nambu-Goto predictions for the 
lowest energy levels. (Recall that the latter are determined by the adjoint 
string tension, which in turn is determined by the ground state energies at
larger $l$.) We see that the observed spectrum bears no relation at all to 
the simple Nambu-Goto spectrum (even at our largest $l$) in marked contrast 
to the remarkable success
of the latter in describing the spectrum of fundamental flux tubes. Even if one
looks at the deviations of the ground state from Nambu-Goto, and fits that
with an inverse power, one finds a power roughly $\propto 1/l^3$ which is
inconsistent with what one expects from the universal terms in an effective
string action
\cite{OAZK,OA}.
We have seen in Figs.\ref{fig_adjtotvac_n4},\ref{fig_adjgsvac_n4} 
that the vacuum overlaps are very 
small, and there is no sign of a glueball contribution to the correlator, so
the decays and mixings that would be suppressed at large-$N$ appear to be
negligible here. It may be that it is the mixings and decays involving a pair of 
(anti)fundamental flux tubes that lead to the complexity of our observed
adjoint spectrum. In the large-$N$ limit the spectrum should become that
of two non-interacting fundamental flux tubes, each with its own excitation
spectrum based on a string with tension $\sigma_f$. This will be
different from the spectrum of a single string with a tension
$\lim_{N\to\infty}\sigma_{adj} = 2\sigma_f$, both for the excitation
energy at $O(1/l^3)$ and for the zero-point energy at $O(1/l)$ and,
in particular, because of the sum over the relative transverse momenta of 
the two (anti)fundamental flux tubes.


%
%
\subsection{$p\neq 0$ spectra}
\label{subsection_pneq0}

A striking feature of the calculated flux tube spectra is how closely the 
ground states with non-zero longitudinal momentum, $p$, follow the Nambu-Goto 
spectrum in eqn(\ref{eqn_EnNG}). In the case of $SU(6)$ one sees this not only
for the fundamental flux tube
\cite{AABBMT11}
but also for the flux in higher representations such as, amongst others, 
the $k=2A,2S$, $k=3A,3M,3S$, and adjoint $k=0$
\cite{AAMT13}.
In Fig.\ref{fig_E_gsqall_su2} we show how the energies of the $SU(2)$ ground 
states vary with $l$ for longitudinal momenta $p=0,2\pi/l,4\pi/l,6\pi/l$. 
We also show the Nambu-Goto predictions. We see that even for $SU(2)$ the
agreement with Nambu-Goto is very good, with only very small deviations
at the smallest values of $l$. One can ask whether these deviations
fall  within our systematic errors. In particular, our Nambu-Goto spectrum
is for the continuum theory, but
at the smallest values of $l$ our values of $ap$ are not small, and
one should then worry about $O((ap)^2)$ lattice spacing corrections. 
While we are not able to quantify these corrections, we can obtain 
an estimate of their potential magnitude by replacing the continuum 
momentum by the one which arises in the free field propagator of the
most simply discretised bosonic action
\begin{equation}
(ap)^2 \to 2 - 2 \cos (ap) .
\label{eqn_plattice}
\end{equation}
Doing so we obtain the dashed curves in Fig.\ref{fig_E_gsqall_su2}.
We note that the lattice spacing corrections are very small, but comparable 
in size to the small deviations of the calculated energies from continuum 
Nambu-Goto. So while the deviations may well be real, we cannot claim to
estimate these since they appear to be within our systematic uncertainties.

Given that the close agreement at smaller values of $l$ is not just at large 
$N$ but extends all the way down to $SU(2)$, and also that in the case of $SU(2)$ 
it extends below the 
value of $l$ where the $p=0$ Nambu-Goto ground state becomes tachyonic, one needs 
to ask what is the significance of this apparent agreement with Nambu-Goto. 
The first thing to note is that much of the gap between $p\neq 0$ and $p=0$ 
at small $l$ arises from the contribution of the momentum, $p=2\pi n/l$ which 
becomes large as $l$ decreases, even for $n=1$, and will dominate the energy. 
We see this in Fig.\ref{fig_E_gsqall_fits_su2}, where each dashed line is 
simply the energy of a massless particle of momentum $p$, i.e. $E=p=2\pi n/l$.
Of course in the confining phase we believe that we have a background flux tube. 
If we take the simplest possible model, which is that the phonons on the world 
sheet are completely non-interacting (see below), then the energy of the ground 
state of momentum $p$  will be $E=p=2\pi n/l$ plus that of the background 
flux tube, $\sigma l$, i.e.
\begin{equation}
E_{free,gs}^2(l;p) = \left(\sigma l + \frac{2\pi n}{l}\right)^2 
=(\sigma l)^2 + 4\pi\sigma n + \left(\frac{2\pi n}{l}\right)^2.
\label{eqn_Epfree}
\end{equation}
This is plotted as the solid lines in Fig.\ref{fig_E_gsqall_fits_su2} and 
by eye it agrees as well as Nambu-Goto did in Fig.\ref{fig_E_gsqall_su2}.
This should be no surprise: the Nambu-Goto energy for this case where all
the phonons are left movers (or all right movers) is
\begin{equation}
E_{NG,gs}^2(l;p) = (\sigma l)^2 + 4\pi\sigma \left(n - \frac{1}{12}\right) 
+ \left(\frac{2\pi n}{l}\right)^2 ,
\label{eqn_EpNG}
\end{equation}
which differs from $E_{free,gs}^2(l;p)$ by an amount that is invisibly small
in Figs.\ref{fig_E_gsqall_su2},\ref{fig_E_gsqall_fits_su2}. Indeed
the subenergy of any two  left-moving massless phonons is zero so they do
not interact since they necessarily have derivative interactions as 
befits Goldstones (arising from the spontaneous breaking of bulk translation 
invariance by the background flux tube). However there are always extra phonons
winding around the $l$-torus, and interacting with these gives rise to
the small extra term in eqn(\ref{eqn_EpNG}), as shown in
\cite{SD_short}
using the Thermodynamic Bethe Ansatz.
(It can also be thought of as arising from the zero-point energies of the
phonon oscillators, at least to a first approximation.) 

So we see that most of the gap between the $E_{gs}^2(l;p)$ for two different $p$ 
is due to the energy of the massless phonons in the approximation that they are 
free. To discriminate betwen the `free' and `NG' models we have to take a 
microscope to our data, and this we do for $p=2\pi/l$ (our most accurate 
$p\neq 0$ calculations) in Fig.\ref{fig_DEmodelgs_k1q1P-_su2} for $SU(2)$, and 
in Fig.\ref{fig_DEmodelgs_k1q1P-_su6} for $SU(6)$. From these figures it is clear 
that the very small extra piece in the Nambu-Goto energy is indeed necessary 
(and sufficient) to obtain excellent agreement with our calculated values. 

It is interesting to ask to what extent our spectrum constrains the mass $\mu$ 
of the momentum carrying excitation to be zero, as assumed above. To a good
approximation a non-zero mass $\mu$ will simply shift the value of the 
quantity $l(E-E_{model})/2\pi$  plotted in 
Figs.\ref{fig_DEmodelgs_k1q1P-_su2},\ref{fig_DEmodelgs_k1q1P-_su6}
by  $\mu^2l^2/8\pi^2 = (\mu^2/\sigma)(l\surd\sigma)^2/8\pi^2$, and this will
be very small at small $l$. Nonetheless the agreement with Nambu-Goto is
so good that the constraint turns out to be quite strong: we find
$\mu^2/\sigma \ll 1$ for all $l\surd\sigma \gtrsim 1.5$. Since the bulk
mass gap is $m_g/\surd\sigma \sim 4$ (although some screening masses
are smaller by perhaps a factor of $\sim 2$), this result is telling us 
that the string excitations are indeed massless, as one expects.

A closed flux tube of length $l\surd\sigma \sim 2$ is almost as wide as it
is long. (Assuming a conventional intrinisic width of $\sim 1/\surd\sigma$.)
So, naively, it looks like a fat blob rather than an ideal string. That its
excitations should be given so accurately by the massless phonons 
characteristic of a thin string is therefore a surprise that is telling
us something interesting about the dynamics of flux tubes.

\subsection{continuum and $N\to\infty$}
\label{subsection_contN}

In our complementary study of the glueball spectrum
\cite{AAMT16}
we have also calculated the flux tube spectrum. These calculations are for 
a large range of $N$, so as to enable an $N\to \infty$ extrapolation of the
glueball masses, as well as for a range of lattice spacings so as to enable
a continuum $a\to 0$ extrapolation. However, being aimed primarily at the
glueball spectrum, they are typically performed at each value of $N$ and
$a(\beta)$ for only one or two values of $l$. These values of $l$  are simply
chosen to be large enough for the finite volume corrections to glueball masses
to be negligible and they are not exactly the same, either for fixed $N$ and
varying $a$, or for different $N$. Since the flux tube spectrum depends quite
strongly on $l$, this means that any comparison based on the results of
\cite{AAMT16}
will, in the present context, have some significant systematic uncertainties.
To minimise these, we remove the leading $l$-dependence by plotting the
excited energies in a form that, in the Nambu-Goto approximation,
exposes the phonon excitation energy with the dependence on $l$ completely
factored out. For the lightest states with $p=0$ and $n=N_L+N_R$, we plot
the quantity
\begin{equation}
  \frac{E_n^2(l) - E_0^2(l)}{4\pi\sigma}
\stackrel{NG}{=} n ,
\label{eqn_DE2NG}
\end{equation}
where the equality holds if the calculated values coincide
with the Nambu-Goto model. This provides a convenient way to focus on
the deviations from Nambu-Goto. While these are $l$-dependent (see
for example Fig.\ref{fig_Ek1_q0Pall_su6}) this dependence is quite weak
once $l$ is modestly large, as it will be here.

In Fig.\ref{fig_DENGk1_cont_n8} we plot $(E_n^2 - E_0^2)/4\pi\sigma$
for the lightest few $p=0$ (fundamental) flux tube states in $SU(8)$, 
as a function of $a^2\sigma$. The values of $l$ are not equal but they lie 
in a narrow window, $l\surd\sigma \in (3.78,3.91)$, and as we can see
from  Figs.\ref{fig_Ek1_q0Pall_su6},\ref{fig_Ek1_q0Pall_su8} any variation
within this window should be small for these states. What we see in
Fig.\ref{fig_DENGk1_cont_n8} is that as soon as we are away from the 
coarsest lattice spacing, say $a^2\sigma \lesssim 0.04$, any discretisation
errors are not noticeable. This is reassuring: our study of the
$l$-dependence in this paper was carried out for a much smaller lattice 
spacing, namely $a^2\sigma \simeq 0.0075$. 

In Fig.\ref{fig_DENGk1_suNc} we attempt a comparison of the spectra 
obtained for $N=2,3,4,6,8,12$. The values of $l$ now range over a somewhat 
wider window,  $l\surd\sigma \in (3.55,3.85)$. The lattice spacing
also varies, $a\surd\sigma \in (0.069,0.074)$, but this should be harmless
given the discussion of the previous paragraph. In any case we observe that 
at least for $N\geq 6$ any dependence on $N$ is too small to see.
This complements the comparison between $SU(6)$ 
and $SU(8)$ in Figs.\ref{fig_Ek1_q0Pall_su6},\ref{fig_Ek1_q0Pall_su8}, and
reinforces our earlier assertion in
\cite{AABBMT11}
that the spectrum of $SU(6)$ is a good approximation to the $SU(\infty)$
spectrum.

\section{Conclusions}
\label{section_conclusion} 

In this paper we have extended our earlier calculations of the spectrum
of closed confining flux tubes that wind around a spatial torus.
A major motivation has been to address questions brought into focus by the 
remarkable recent progress in our theoretical understanding of the dynamics 
of flux tubes
\cite{OAZK,OA,SD_long,SD_short,SD_massive}.

The  analysis in
\cite{OAZK,SD_long}
of the universal terms in any effective string action predicts that the
first non-universal correction to the ground state energy should be
$O(1/l^{\gamma\geq 7})$. While our earlier work
\cite{AABBMT11}
was only able to show that $\gamma \geq 5$, in the present paper our calculations 
of fundamental flux tubes in $SU(8)$ and in $SU(4)$, and of the $k=2$ flux tube 
in $SU(4)$, unambiguously support the conclusion that $\gamma \geq 7$. 
(See Figs.\ref{fig_pvalue1-4_su8},\ref{fig_pvalue1-5_su8},\ref{fig_pvalue_su6},\ref{fig_pvalue_su4},\ref{fig_pvalue_su4k2A}.) 
This provides further evidence that the theoretical framework of an effective 
string action is appropriate for describing the stable confining flux tubes 
of $SU(N)$ gauge theories, even when $N$ is far from being large.

The calculations in
\cite{SD_short,SD_massive},
using the Thermodynamic Bethe Ansatz to relate the observed energies
to world-sheeet scattering phase shifts, provide convincing evidence that
the first excited state of the $k=2$ flux tube spectrum, as calculated in
\cite{AAMT13}, 
is indeed a massive (resonant) mode on the world sheet, as surmised, 
inconclusively, in 
\cite{AAMT13}.
In the present paper we calculate the spectra of $k=2$ flux tubes in $SU(4)$ and 
$SU(6)$ and $k=3$ flux tubes in $SU(6)$, as shown in 
Figs.\ref{fig_Ek2A_q0P+_su6},\ref{fig_Ek2A_q0P+_su4},\ref{fig_Ek3A_q0P+_su6}. 
Comparing the $k=2$ massive modes in $SU(4)$ and $SU(6)$ we find 
that the scaling is much better when it is expressed in units of the bulk mass 
gap, i.e. the lightest scalar glueball, than when it is expressed in terms of 
the world-sheet ($k=2$) tension, as we can see from 
Figs.\ref{fig_DE1E0k2A_sigk2A_n4-16},\ref{fig_DE1E0k2A_m0pp_n4-16}. 
That is to say, the origins of this massive mode lie in the bulk physics, 
as indeed one would expect if it reflected the binding of two 
fundamental flux tubes into the $k=2$ flux tube.

In addition to the above we also investigated the remarkable suppression of 
screening of flux tubes in representations higher than the fundamental, 
already pointed out in
\cite{AAMT13,sigma_k2,AABBMT}.
Our main improvement is to go from the earlier basis of operators that was not 
orthogonal, to the orthonormal basis provided by our variational procedure. 
(So that our operators correspond to our best estimates for the flux tube energy 
eigenoperators.) This enables us to calculate quantities such as the total
overlap between the spaces spanned by two bases in different representations.
Since these bases typically contain most of the low-lying eigenoperators,
this provides a quantitative measure of the physically interesting screening
of flux tubes in one representation to the other. As an example we showed in 
Fig.\ref{fig_k2Atotk2S_n4} the overlap in $SU(4)$ between the $k=2$ symmetric 
and antisymmetric bases. The overlap decreases very rapidly with increasing
flux tube length, and is only significant for operators that are so highly
smeared that they wrap strongly around the whole spatial volume. (Why
this should be so remains to be understood.) In
Figs.\ref{fig_k2Agsk2S_n4},\ref{fig_k2Sgsk2A_n4} we showed the overlap
of the $k=2A,\,2S$ ground states onto the $2S$ and $2A$ bases respectively.
Again we see a strong suppression of screening. Our study of the overlap
of the adjoint flux tube onto the (singlet) vacuum in $SU(2)$ and $SU(4)$
showed the same features. (See Figs.\ref{fig_adjtotvac_n2}--\ref{fig_adjgsvac_n2}.)

Amongst other calculations in this paper, we performed a comparison of flux 
tube spectra for $SU(N)$ groups ranging from $SU(2)$ to $SU(8)$, as displayed in 
Figs.\ref{fig_Ek1_q0Pall_su2}--\ref{fig_Ek1_q0Pall_su8}. We saw that even for 
$SU(2)$ the spectrum converges with remarkable speed to the Nambu-Goto predictions 
as $l$ increases. In marked contrast, the spectrum of the adjoint flux tube
in $SU(4)$, shown in Fig.\ref{fig_Ek0_q0P++_su4}, bears so little relation
to the Nambu-Goto spectrum that from this spectrum alone we would not
imagine that Nambu-Goto might have something to do with real-world flux
tubes. This despite the fact that a well defined ground state for the adjoint 
flux tube appears to exist in $SU(4)$, as we infer from 
Figs.\ref{fig_Eeffk0gs_n4},\ref{fig_Egsk0bl456b_su4}.
(Perhaps we are seeing in Fig.\ref{fig_Ek0_q0P++_su4} a mix of genuine 
adjoint states and states composed of a pair of (anti)fundamental flux 
tubes with equal and opposite momenta.) We also looked more critically at 
the remarkable apparent agreement between the calculated $p\neq 0$ ground 
state energies and the Nambu-Goto predictions, which persists 
to the smallest values of $l$, even for $SU(2)$, as we see in
Fig.\ref{fig_E_gsqall_su2}. In fact, as we saw in
Fig.\ref{fig_E_gsqall_fits_su2}, the same level of visual agreement
is obtained with a simple model of free massless phonons on a background 
flux tube. And indeed at the smallest values of $l$ it is simply
the energy due to the momentum that dominates the energy gaps.
If however we apply a microscope to the comparisons, as in
Figs.\ref{fig_DEmodelgs_k1q1P-_su2},\ref{fig_DEmodelgs_k1q1P-_su6},
then we see quite clearly that it is only Nambu-Goto that provides a 
truly accurate description of the calculated energies.

Some final comments. The fact that the first correction to the Nambu-Goto 
prediction for the energy spectrum is $\propto 1/(l\surd\sigma)^{\gamma\geq 7}$
\cite{OAZK,OA}
means that the spectrum of closed flux tubes is accurately known for all lengths 
$l$ where the expansion in powers of $1/l^2\sigma$ is convergent. 
The fact that at smaller $l$ the world sheet theory is approximately integrable
\cite{SD_short}
means that we have an accurate description for (most) smaller values of $l$ as well.
Moreover the corresponding theoretical formalism allows us to identify and 
incorporate new world sheet phenomena such as massive modes
\cite{SD_massive}. 
Such massive modes have been identified in both $D=3+1$ fundamental flux tubes
\cite{AABBMT10,SD_massive}
and in $k$-strings
\cite{AAMT13,SD_massive}.
Although these massive modes belong to the world-sheet theory they may well
influence the glueball spectrum of the bulk $SU(N)$ theory. In a picture where
glueballs are based on contractible closed loops of fundamental flux (as at least 
some of them surely are) the massive modes will produce corresponding excitation 
gaps in in the glueball energy spectrum. This could be encoded in simple
flux tube models of glueballs
\cite{NIJP,RJMT}.
Finally we note that the strong suppression of gluon screening implied
by the observation that the flux carried by long flux tubes remains, to a very 
good approximation, in a given $SU(N)$ representation, means that 
an approximation to the theory that neglects screening, e.g.
\cite{Nair}, 
and so is more tractable, may nonetheless form the basis for a good approximation 
to the dynamics of flux tubes, and perhaps even the bulk theory, in $D=2+1$.

\section*{Acknowledgements}

We have benefitted from very useful discussions with, in particular, Ofer Aharony, 
Sergei Dubovsky, Raphael Flauger, Victor Gorbenko and Zohar Komargodski, as well as 
with other participants at both the recent {\it Flux Tubes} conference held at the 
Perimeter Institute (in May 2015), the earlier {\it Confining Flux Tubes and Strings}
conference held at ECT, Trento (in July 2010), and the much earlier {\it QCD and Strings}
workshop at the KITP, UCSB (in 2004). We are grateful to these institutions
for hosting these very productive meetings. 
AA has been partially supported by an internal program of the University of Cyprus 
under the name of BARYONS. In addition, AA acknowledges the hospitality of the 
Cyprus Institute where part of this work was carried out. 
MT acknowledges partial support under STFC grant ST/L000474/1.
The numerical computations were carried out on the computing cluster 
in Oxford Theoretical Physics.

\vspace*{3.0cm}

\begin{appendix}
%
%
%
%
\section{Tables}
\label{section_appendix_results}




\begin{table}[htb]
\begin{center}
  \begin{tabular}{|cc|ccc|c|}\hline
 group    & $\beta$ & $a^2\sigma_{k=1}$ & $a^2\sigma_{k=2A}$ & $a^2\sigma_{k=3A}$ & $l_c\surd\sigma_{k=1}$  \\ \hline
SU(2) & 16.0   & 0.0076416(46) & --           & --           &  0.8953(72) \\
SU(4) & 74.0   & 0.0074480(55) & 0.010098(9)  & --           &  1.0543(37) \\
SU(6) & 171.0  & 0.0073662(37) & 0.01198(3)   & 0.01357(5  ) &  1.0865(41) \\
SU(8) & 306.25 & 0.0073745(43) & 0.01298(4)   & 0.01635(10)  &  1.1000(30) \\  \hline
\end{tabular}
  \caption{
    String tensions, $\sigma_{k}$,  and the critical length, $l_c$, below which the theory is no longer 
    confining, for the  main calculations in this paper.}
\label{table_suNsigmaTc}
\end{center}
\end{table}

\begin{table}[htb]
\begin{center}
  \begin{tabular}{|cc|cccc|cc|}\hline
\multicolumn{8}{|c|}{ $SU(8) \ \  ; \ \  aE(l,p=0)$ } \\ \hline
$l/a$ & $l_\perp\times l_t$ & \multicolumn{4}{|c|}{ $P=+$ }& \multicolumn{2}{|c|}{ $P=-$ } \\ \hline
14  & $100\times 200$  & 0.05187(14) &  0.3839(62) &            &            & 0.550(16)  & 0.740(21)  \\
15  & $100\times 200$  & 0.06576(14) &  0.3985(86) &            &            & 0.572(20)  & 0.766(25) \\
16  & $100\times 200$  & 0.07810(11) &  0.3967(28) & 0.524(19)  & 0.644(18)  & 0.5433(43) & 0.675(11) \\
17  & $100\times 200$  & 0.08916(24) &  0.4031(20) & 0.5433(85) & 0.645(12)  & 0.5560(46) & 0.654(11) \\
18  & $100\times 200$  & 0.09950(17) &  0.4135(13) & 0.5673(40) & 0.6581(58) & 0.5624(32) & 0.619(13)  \\
19  & $100\times 200$  & 0.10887(18) &  0.4139(29) & 0.5803(47) & 0.6564(68) & 0.5637(48) & 0.648(14) \\
20  &  $70\times 120$  & 0.11868(27) &  0.4214(16) & 0.5888(41) & 0.6333(78) & 0.5741(41) & 0.6396(94) \\
21  &  $70\times 120$  & 0.12679(55) &  0.4220(28) & 0.5914(45) & 0.6520(75) & 0.5674(63) & 0.6399(88) \\
22  &  $70\times 120$  & 0.13653(47) &  0.4237(28) & 0.5958(71) & 0.6355(75) & 0.5811(42) & 0.6358(77) \\
23  &  $70\times 120$  & 0.14509(46) &  0.4245(44) & 0.5948(57) & 0.6538(54) & 0.5855(77) & 0.657(11) \\
24  &  $48\times 60$   & 0.15350(70) &  0.4374(40) & 0.605(12)  & 0.635(12)  & 0.5862(62) & 0.639(10) \\
25  &  $48\times 60$   & 0.15966(108)&  0.4403(49) & 0.6154(57) & 0.6518(93) & 0.5936(89) & 0.657(11),\\
26  &  $48\times 60$   & 0.17039(62) &  0.4408(48) & 0.6190(72) & 0.6589(78) & 0.5928(49) & 0.659(15) \\
27  &  $48\times 60$   & 0.17836(71) &  0.4477(47) & 0.6078(93) & 0.6554(62) & 0.6107(76) & 0.656(12) \\
28  &  $48\times 60$   & 0.18730(73) &  0.4559(37) & 0.604(11)  & 0.6513(71) & 0.5986(99) & 0.66195(85) \\
29  &  $48\times 60$   & 0.19425(71) &  0.4614(21) & 0.603(14)  & 0.650(15)  & 0.6082(95) & 0.684(13) \\
30  &  $48\times 60$   & 0.20142(78) &  0.4644(37) & 0.625(11)  & 0.629(14)  & 0.6263(74) & 0.688(18) \\
31  &  $40\times 48$   & 0.20889(80) &  0.4664(45) & 0.609(13)  & 0.6687(62) & 0.621(12)  & 0.70015(87) \\
32  &  $40\times 48$   & 0.21944(80) &  0.4737(40) & 0.6370(79) & 0.6610(59) & 0.6190(83) & 0.64489(84) \\
64  &  $64\times 64$   & 0.46461(114)&  0.6346(53) & 0.7505(59) & 0.7580(64) & 0.7621(54) & 0.766(13) \\ \hline
\end{tabular}
\caption{The energies, $E(l,p)$, of the lightest flux tube states with 
length $l$, parity  $P=P_t=\pm$ and longitudinal momentum $p=0$. 
For SU(8) at $\beta=306.25$.}
\label{table_su8Q0}
\end{center}
\end{table}

\begin{table}[h]
\begin{center}
  \begin{tabular}{|c|cccc|cc|}\hline
\multicolumn{7}{|c|}{ $SU(6) \ \  ; \ \  aE(l,p=0)$ } \\ \hline
$l/a$ & \multicolumn{4}{|c|}{ $P=+$ } & \multicolumn{2}{|c|}{ $P=-$ } \\ \hline
13  & 0.03301(24)  & 0.255(15)  &            &            &            &   \\
14  & 0.05234(29)  & 0.355(13)  &            &            &            &   \\
15  & 0.06581(18)  & 0.3981(72) &            &            &            &   \\
16  & 0.07783(11)  & 0.3905(54) & 0.419(30)  & 0.601(13)  & 0.486(13)  &  0.597(14) \\
17  & 0.08890(26)  & 0.4034(39) & 0.513(15)  & 0.607(14)  & 0.551(10)  &  0.632(15)  \\
18  & 0.09894(25)  & 0.4084(27) & 0.560(13)  & 0.630(10)  & 0.542(11)  &  0.614(11) \\
19  & 0.10840(26)  & 0.4163(37) & 0.573(10)  & 0.635(16)  & 0.5571(78) &  0.649(15)  \\
20  & 0.11805(25)  & 0.4220(12) & 0.5802(55) & 0.609(16)  & 0.5578(75) &  0.640(16) \\
21  & 0.12739(42)  & 0.4212(44) & 0.5781(75) & 0.618(15)  & 0.5802(82) &  0.644(12) \\
22  & 0.13660(68)  & 0.4327(22) & 0.6003(87) & 0.632(11)  & 0.5798(84) &  0.646(12) \\
23  & 0.14571(58)  & 0.4296(48) & 0.6021(87) & 0.6469(70) & 0.5999(81) &  0.6490(92)  \\
24  & 0.15308(21)  & 0.4367(27) & 0.5938(65) & 0.6304(54) & 0.5963(54) &  0.6364(68) \\
25  & 0.16177(41)  & 0.4410(24) & 0.6066(70) & 0.6470(53) & 0.5837(90) &  0.6569(66) \\
26  & 0.17054(41)  & 0.4491(33) & 0.6025(73) & 0.6454(80) & 0.6048(89) &  0.6665(65)  \\
27  & 0.17806(46)  & 0.4521(17) & 0.5952(95) & 0.629(10)  & 0.6035(42) &  0.654(14)  \\
28  & 0.18721(63)  & 0.4541(23) & 0.6164(61) & 0.6455(90) & 0.6111(39) &  0.6676(61) \\
29  & 0.19507(49)  & 0.4550(40) & 0.6112(94) & 0.6508(55) & 0.6159(53) &  0.6639(65)  \\
30  & 0.20371(56)  & 0.4675(40) & 0.6134(93) & 0.648(11)  & 0.6164(89) &  0.679(10)  \\
31  & 0.21005(46)  & 0.4622(55) & 0.6181(62) & 0.6631(68) & 0.6321(98) &  0.672(14)   \\
32  & 0.2177(9)    & 0.4724(42) & 0.6021(94) & 0.631(12)  & 0.604(10)  &  0.674(15)  \\
36  & 0.2494(11)   & 0.4921(20) & 0.6290(49) & 0.6475(57) & 0.6384(80) &  0.6683(62)   \\
40  & 0.2820(10)   & 0.5064(30) & 0.6464(57) & 0.6655(57) & 0.6573(52) &  0.6761(52) \\
44  & 0.3114(11)   & 0.5292(34) & 0.652(11)  & 0.6781(73) & 0.6885(51) &  0.6899(50)   \\
48  & 0.3424(10)   & 0.5459(36) & 0.678(11)  & 0.6985(60) & 0.6964(61) &  0.718(12)   \\
52  & 0.3731(9)    & 0.5621(37) & 0.7072(91) & 0.7196(57) & 0.7199(72) &  0.725(14)  \\
56  & 0.4060(11)   & 0.5825(55) & 0.720(11)  & 0.7376(80) & 0.7394(57) &  0.731(16)  \\
60  & 0.4339(10)   & 0.6089(68) & 0.733(11)  & 0.7505(78) & 0.7462(74) &  0.7506(72)  \\
64  & 0.4626(16)   & 0.6322(38) & 0.7576(81) & 0.7632(82) & 0.7605(53) &  0.789(19)  \\ \hline
  \end{tabular}
\caption{The energies, $E(l,p)$, of the lightest fundamental flux tube states with 
length $l$, parity $P=P_t=\pm$  and longitudinal momentum $p=0$. 
For SU(6) at $\beta=171.0$.}
\label{table_su6Q0}
\end{center}
\end{table}

\begin{table}[h]
\begin{center}
  \begin{tabular}{|cc|cccc|c|c|}\hline
\multicolumn{8}{|c|}{ $SU(4) \ \  ; \ \  aE(l,p=0)$ } \\ \hline
$l/a$ & $l_\perp\times l_t$ & \multicolumn{4}{|c|}{ $P=+,+$ } & $P=-,-$ & $P=-,+$ \\ \hline
13  & $100\times 200$  & 0.03291(60) &            &            &            & 0.6362(43) & 0.228(22)  \\
14  &  $80\times 160$  & 0.05230(61) & 0.214(8)   &            &            & 0.6398(21) & 0.281(30) \\ 
15  &  $80\times 160$  & 0.06609(46) & 0.256(20)  &            &            & 0.6292(40) & 0.336(8) \\ 
18  &  $70\times 120$  & 0.10067(34) & 0.324(20)  &            &            & 0.6276(39) & 0.448(8) \\ 
22  &  $60\times  80$  & 0.13740(46) & 0.4285(21) & 0.470(12)  & 0.6044(42) & 0.6247(64) & 0.5911(22) \\ 
26  &  $52\times  64$  & 0.17213(44) & 0.4504(29) & 0.532(7)   & 0.6199(44) & 0.6341(44) & 0.6040(38) \\
30  &  $48\times  64$  & 0.20607(47) & 0.4707(16) & 0.575(13)  & 0.6333(24) & 0.6493(23) & 0.6253(50)  \\ 
34  &  $48\times  64$  & 0.23684(57) & 0.4866(17) & 0.5980(76) & 0.6517(14) & 0.6505(47) & 0.6405(24) \\ 
38  &  $48\times  64$  & 0.26946(50) & 0.5074(9)  & 0.6381(22) & 0.6678(13) & 0.6650(47) & 0.6583(22) \\ 
44  &  $52\times  64$  & 0.31510(62) & 0.5363(10) & 0.6551(43) & 0.6887(24) & 0.6981(31) & 0.6800(25) \\ 
52  &  $52\times  64$  & 0.3774(7)   & 0.5724(22) & 0.7037(51) & 0.7092(33) & 0.7246(49) & 0.7184(20) \\ 
58  &  $58\times  64$  & 0.4226(5)   & 0.6059(17) & 0.7328(38) & 0.7429(44) & 0.7451(44) & 0.7485(21) \\ \hline 
\end{tabular}
\caption{The energies, $E(l,p)$, of the lightest fundamental flux tube states with 
length $l$, parity $P= \{P_t,P_\parallel\}$ and longitudinal momentum $p=0$. 
For SU(4) at $\beta=74.0$.}
\label{table_su4Q0}
\end{center}
\end{table}

\begin{table}[h]
\begin{center}
  \begin{tabular}{|cc|cccc|c|c|}\hline
\multicolumn{8}{|c|}{ $SU(2) \ \  ; \ \  aE(l,p=0)$ } \\ \hline
$l/a$ & $l_\perp\times l_t$ & \multicolumn{4}{|c|}{ $P=+,+$ } & $P=-,-$ & $P=-,+$ \\ \hline
11  & $100\times 160$  & 0.01898(20)  & 0.1053(53)  & 0.167(16)  & 0.294(12)  & 0.8422(13) & 0.3075(48) \\
14  &  $80\times 140$  & 0.06128(16)  & 0.2593(29)  & 0.355(17)  & 0.377(42)  & 0.7574(64) & 0.4372(38) \\
18  &  $70\times 120$  & 0.10504(16)  & 0.3997(51)  & 0.5051(35) & 0.507(10)  & 0.7201(13) & 0.5728(57) \\
22  &  $60\times  80$  & 0.14180(29)  & 0.5045(11)  & 0.499(10)  & 0.6529(45) & 0.7013(12) & 0.6430(22) \\
26  &  $52\times  64$  & 0.17750(24)  & 0.5028(11)  & 0.5914(46) & 0.7000(40) & 0.6942(18) & 0.6881(31)  \\
30  &  $48\times  64$  & 0.21075(37)  & 0.5045(18)  & 0.6376(56) & 0.7014(62) & 0.6812(41) & 0.7185(32) \\
34  &  $48\times  64$  & 0.24382(43)  & 0.5155(8)   & 0.6445(91) & 0.7361(28) & 0.7000(17) & 0.7284(16) \\
38  &  $48\times  64$  & 0.27612(60)  & 0.5287(13)  & 0.6820(24) & 0.7482(56) & 0.7027(31) & 0.7281(44) \\
44  &  $44\times  54$  & 0.3242(9)    & 0.5537(13)  & 0.6866(69) & 0.7541(73) & 0.7210(38) & 0.7296(32) \\
46  &  $46\times  52$  & 0.34058(60)  & 0.5589(12)  & 0.7103(24) & 0.7657(34) & 0.7328(26) & 0.7371(28) \\
58  &  $58\times  48$  & 0.43466(74)  & 0.6237(21)  & 0.7686(24) & 0.7964(38) & 0.7793(65) & 0.7375(37) \\
68  &  $68\times  48$  & 0.5140(12)   & 0.6848(18)  & 0.8252(27) & 0.8476(44) & 0.8270(92) & 0.771(24) \\ \hline
\end{tabular}
\caption{The energies, $E(l,p)$, of the lightest fundamental flux tube states with 
length $l$, parity  $P= \{P_t,P_\parallel\}$ and longitudinal momentum $p=0$. 
For SU(2) at $\beta=16.0$.}
\label{table_su2Q0}
\end{center}
\end{table}

\begin{table}[h]
\begin{center}
  \begin{tabular}{|c|ll|c|ll|ll|}\hline
\multicolumn{8}{|c|}{ $aE_{k}(l,p=0)$ } \\ \hline
\multicolumn{3}{|c|}{$SU(4)$} & \multicolumn{5}{|c|}{$SU(6)$}\\ \hline
$l/a$ & \multicolumn{2}{|c|}{$k=2A$} & $l/a$ & \multicolumn{2}{|c|}{$k=2A$}& \multicolumn{2}{|c|}{$k=3A$}\\ \hline
13  &            &            & 16 & 0.1460(14) & 0.2778(168) & 0.1742(11) & 0.3962(75) \\
14  & 0.0727(20) & 0.144(12)  & 20 & 0.2088(17) & 0.3794(121) & 0.2433(21) & 0.4627(99) \\
15  &            &            & 24 & 0.2649(23) & 0.4740(80)  & 0.3020(32) & 0.5178(72) \\
18  & 0.1463(9)  & 0.264(11)  & 28 & 0.3198(29) & 0.5346(84)  & 0.3569(39) & 0.570(16) \\
22  & 0.1951(8)  & 0.4092(78) & 32 & 0.3633(22) & 0.5672(63)  & 0.4198(53) & 0.613(12) \\
26  & 0.2409(7)  & 0.4570(85) & 36 & 0.4192(25) & 0.6131(77)  & 0.4762(50) & 0.667(13) \\
30  & 0.2859(9)  & 0.5081(38) & 40 & 0.4615(42) & 0.6551(106) & 0.5259(67) & 0.732(15)  \\
34  & 0.3258(9)  & 0.5484(13) & 44 & 0.5144(50) & 0.7158(133) & 0.5806(74) & 0.789(13)  \\
38  & 0.3700(14) & 0.5891(20) & 48 & 0.5624(40) & 0.7508(142) & 0.6405(79) & 0.840(11) \\
44  & 0.4346(10) & 0.6483(17) & 52 & 0.6183(60) & 0.8019(113) & 0.7015(83) & 0.893(17) \\
52  & 0.5123(14) & 0.7075(26) & 64 & 0.7661(109) & 0.9448(61) & 0.863(14)  & 1.003(41) \\
58  & 0.5769(13) & 0.7646(26) &  &  &  &  &  \\
60  & 0.5974(17) & 0.7817(20) &  &  &  &  &  \\ \hline
\end{tabular}
  \caption{The energies of the ground  and first excited  states
    of a flux tube of length $l$ and momentum $p=0$ carrying flux $k=2A$ and $k=3A$. 
    For $SU(4)$ at $\beta=74$ and $SU(6)$ at $\beta=171$ as shown.} 
\label{table_su4su6Ek2A3A}
\end{center}
\end{table}

\end{appendix}

\clearpage

%
%

\begin{figure}[htb]
\begin	{center}
\leavevmode
\input	{plot_DENGpow9_gsq0k1_n8f.tex}
\end	{center}
\caption{Best fits to SU(8) ground state energy with Nambu-Goto plus a $O(1/l^9)$
correction. Vertical line indicates the deconfining transition.}
\label{fig_DENGpow9_su8}
\end{figure}

\begin{figure}[htb]
\begin	{center}
\leavevmode
\input	{plot_powfit1-4_n8.tex}
\end	{center}
\caption{Best fits to SU(8) ground state using Nambu-Goto with a $O(1/l^\gamma)$
correction for various $\gamma$: $p$-value for $l\in [14,17]$ for three mass 
fits as explained in text.}
\label{fig_pvalue1-4_su8}
\end{figure}

\begin{figure}[htb]
\begin	{center}
\leavevmode
\input	{plot_powfit1-5_n8.tex}
\end	{center}
\caption{Best fits to SU(8) ground state using Nambu-Goto with a $O(1/l^\gamma)$
correction for various $\gamma$: $p$-value for $l\in [14,18]$ for three mass 
fits as explained in text.}
\label{fig_pvalue1-5_su8}
\end{figure}

\begin{figure}[htb]
\begin	{center}
\leavevmode
\input	{plot_DENGpow7_gsq0k1_n6.tex}
\end	{center}
\caption{Best fits to SU(6) ground state energy with Nambu-Goto plus a $O(1/l^7)$
correction. Vertical line indicates the deconfining transition.}
\label{fig_DENGpow7_su6}
\end{figure}

\begin{figure}[htb]
\begin	{center}
\leavevmode
\input	{plot_powfit_n6f.tex}
\end	{center}
\caption{Best fits to SU(6) ground state energy using Nambu-Goto with a $O(1/l^\gamma)$
correction: $p$-value for $l\in [14,17],\, \bullet$, and for
$l\in [14,18], \, \circ$, versus $\gamma$.}
\label{fig_pvalue_su6}
\end{figure}

\begin{figure}[htb]
\begin	{center}
\leavevmode
\input	{plot_DENGpow7_gsq0k1_n4.tex}
\end	{center}
\caption{Best fits to SU(4) $k=1$ ground state energy with Nambu-Goto
  plus a $O(1/l^7)$
correction. Vertical line indicates the deconfining transition.}
\label{fig_DENGpow7_su4}
\end{figure}

\begin{figure}[htb]
\begin	{center}
\leavevmode
\input	{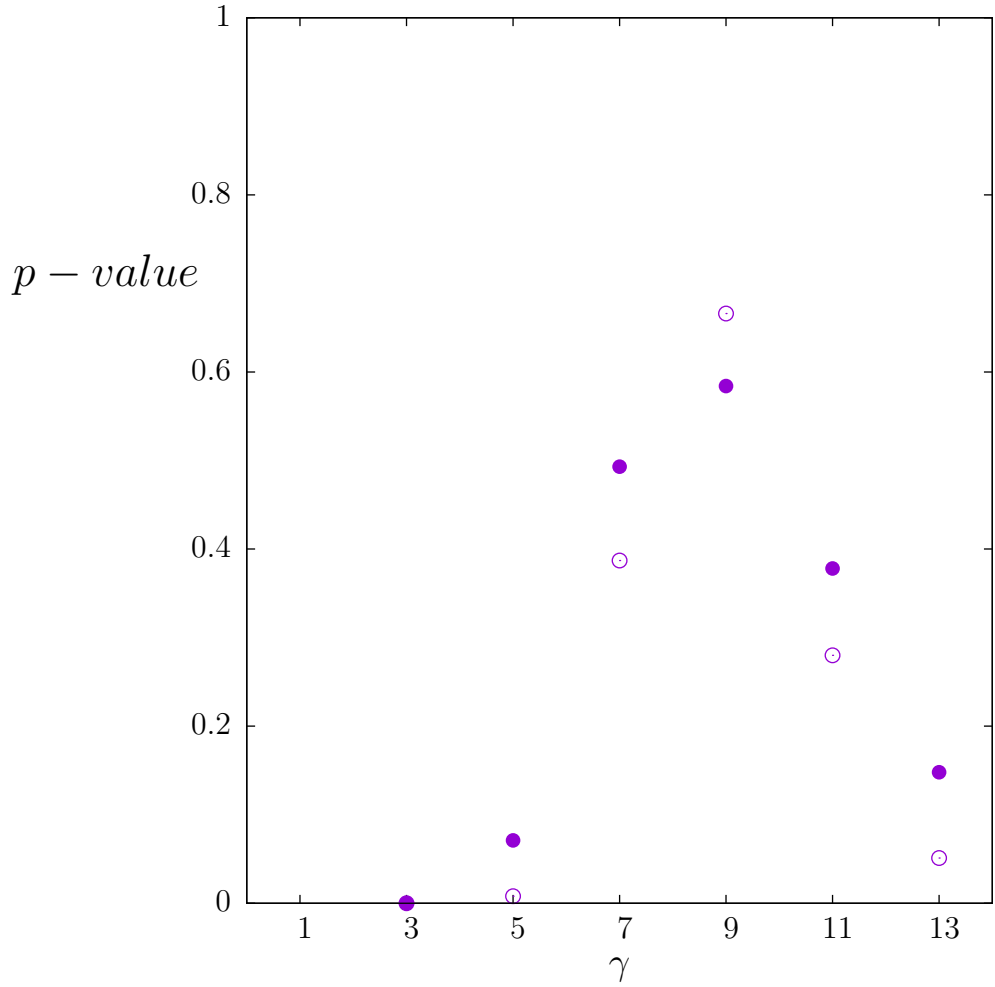}
\end	{center}
\caption{Best fits to SU(4) $k=1$ ground state energy using Nambu-Goto with a $O(1/l^\gamma)$
correction: $p$-value for all $l\in [13,60],\, \bullet$, and for
$l\in [13,18], \, \circ$, versus $\gamma$.}
\label{fig_pvalue_su4}
\end{figure}

\clearpage

\begin{figure}[htb]
\begin	{center}
\input	{plot_DENGpow7_gsq0k2A_n4.tex}
\end	{center}
\caption{Best fits to SU(4) $k=2A$ ground state energy with Nambu-Goto plus a $O(1/l^7)$
correction. Vertical line indicates the deconfining transition.}
\label{fig_DENG_gsk2A_su4}
\end{figure}

\begin{figure}[htb]
\begin	{center}
\leavevmode
\input	{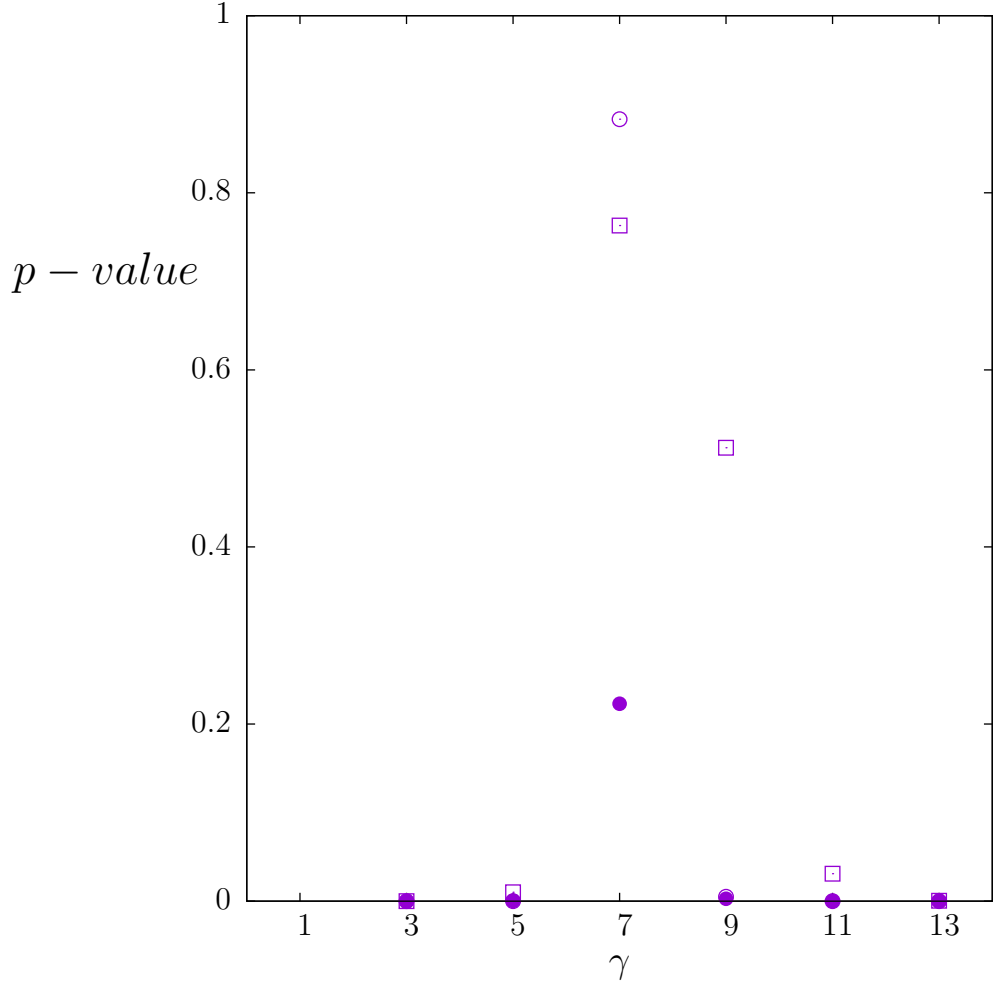}
\end	{center}
\caption{Best fits to SU(4) $k=2A$ ground state energy using Nambu-Goto with a $O(1/l^\gamma)$
correction: $p$-value for all $l\in [13,60],\, \bullet$, and for
$l\in [13,18], \, \circ$, versus $\gamma$. Also fits $l\in [14,18], \, \square$, that exclude
the shortest flux tube.}
\label{fig_pvalue_su4k2A}
\end{figure}

\begin{figure}[htb]
\begin	{center}
\leavevmode
\input	{plot_DENGexp_gsq0k1_n8f.tex}
\end	{center}
\caption{Best fits to SU(8) ground state energy with Nambu-Goto plus a $\propto e^{-\mu l}$
correction. Vertical line indicates the deconfining transition.}
\label{fig_DENGexp_su8}
\end{figure}

\begin{figure}[htb]
\begin	{center}
\leavevmode
\input	{plot_DENGexp_gsq0k2A_n4.tex}
\end	{center}
\caption{Best fits to SU(4) $k=2A$  ground state energy with Nambu-Goto plus
  a $\propto e^{-\mu l}$
correction. Vertical line indicates the deconfining transition.}
\label{fig_DENGk2Aexp_su4}
\end{figure}

%
%

\begin{figure}[htb]
\begin	{center}
\input	{plot_Ek2Aq0P+_n6.tex}
\end	{center}
\caption{SU(6) at $\beta=171$. $k=2A$ spectrum with $p=0$ and parity $P_t=+$,
  Vertical line locates deconfining transition.}
\label{fig_Ek2A_q0P+_su6}
\end{figure}

\begin{figure}[htb]
\begin	{center}
\input	{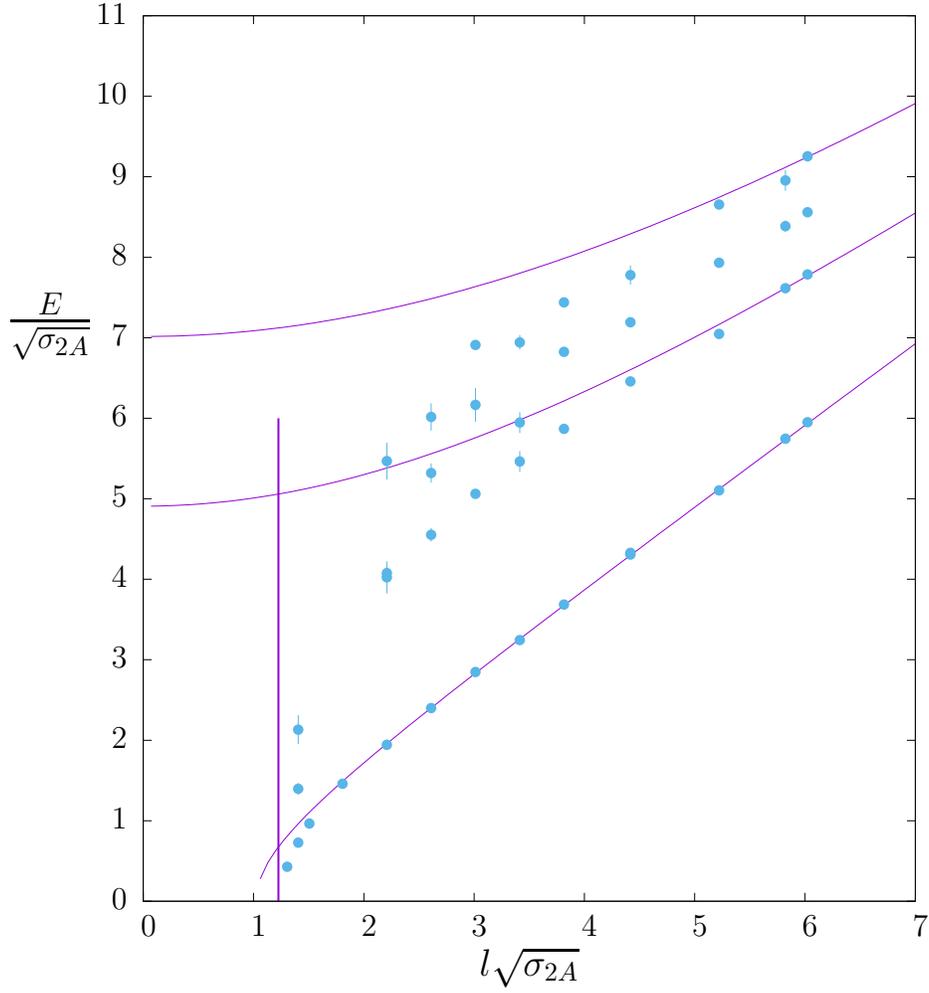}
\end	{center}
\caption{SU(4) at $\beta=74$. Lightest $k=2A$ states with $p=0$ and parities $P_t,P_{\parallel}=+,+$. Vertical line locates deconfining transition.}
\label{fig_Ek2A_q0P+_su4}
\end{figure}

\begin{figure}[htb]
\begin	{center}
\input	{plot_Ek3Aq0P+_n6.tex}
\end	{center}
\caption{SU(6) at $\beta=171$. $k=3A$ spectrum with $p=0$ and parity $P_t=+$,
  Vertical line locates deconfining transition.}
\label{fig_Ek3A_q0P+_su6}
\end{figure}

\begin{figure}[htb]
\begin	{center}
\input	{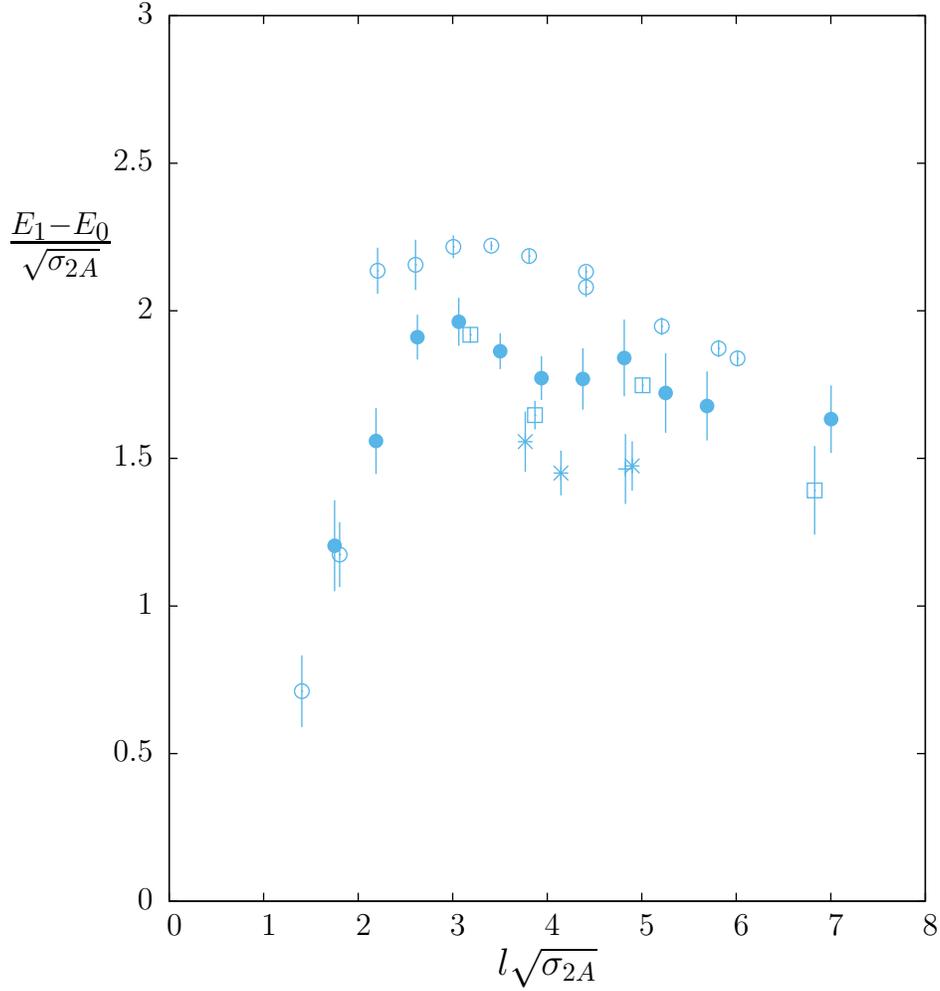}
\end	{center}
\caption{ Difference between ground and first excited energies of a $k=2A$ flux
  tube of length $l$, in units of the $k=2A$ string tension. For $SU(4)$, $\circ$,
  $SU(6)$, $\bullet$, $SU(8)$, $\square$, $SU(12)$, $+$, and $SU(16)$, $\star$.}
\label{fig_DE1E0k2A_sigk2A_n4-16}
\end{figure}

\begin{figure}[htb]
\begin	{center}
\input	{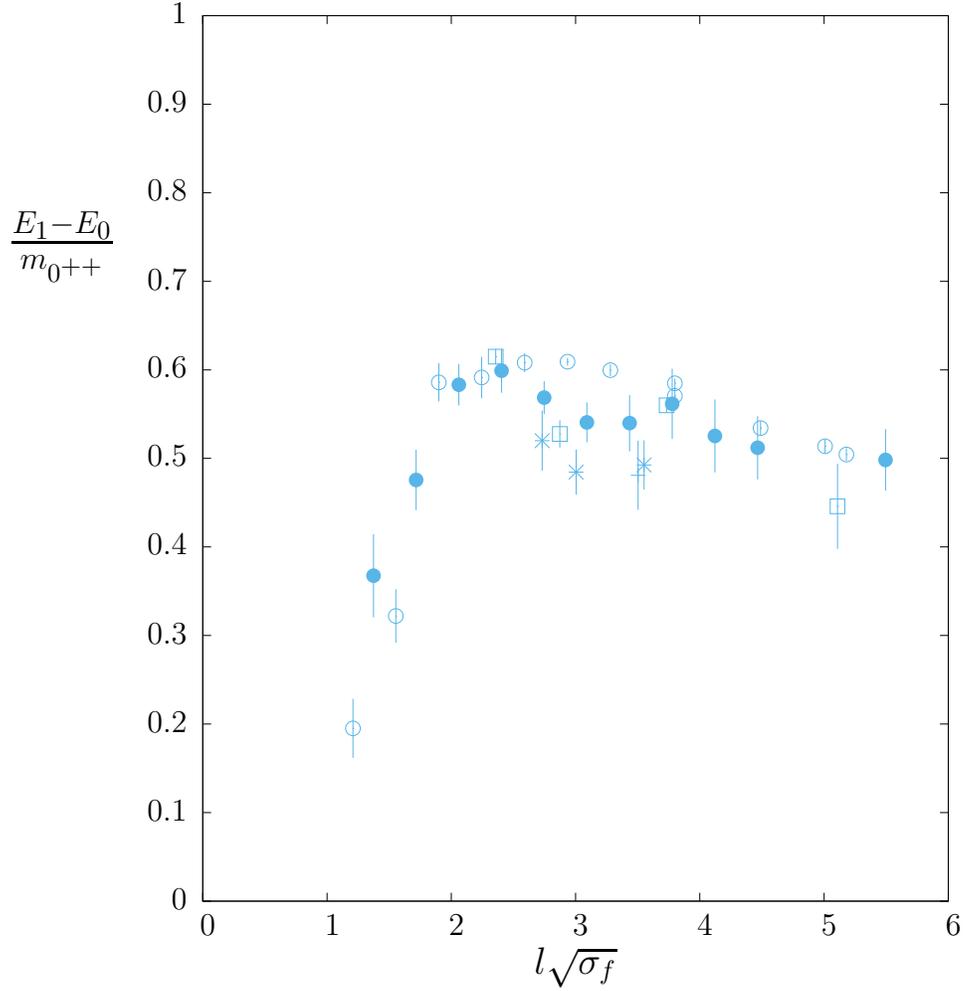}
\end	{center}
\caption{ Difference between ground and first excited energies of a $k=2A$ flux
  tube of length $l$, in units of the $l=\infty$ mass gap. For $SU(4)$, $\circ$,
  $SU(6)$, $\bullet$, $SU(8)$, $\square$, $SU(12)$, $+$, and $SU(16)$, $\star$.}
\label{fig_DE1E0k2A_m0pp_n4-16}
\end{figure}

\begin{figure}[htb]
\begin	{center}
\input	{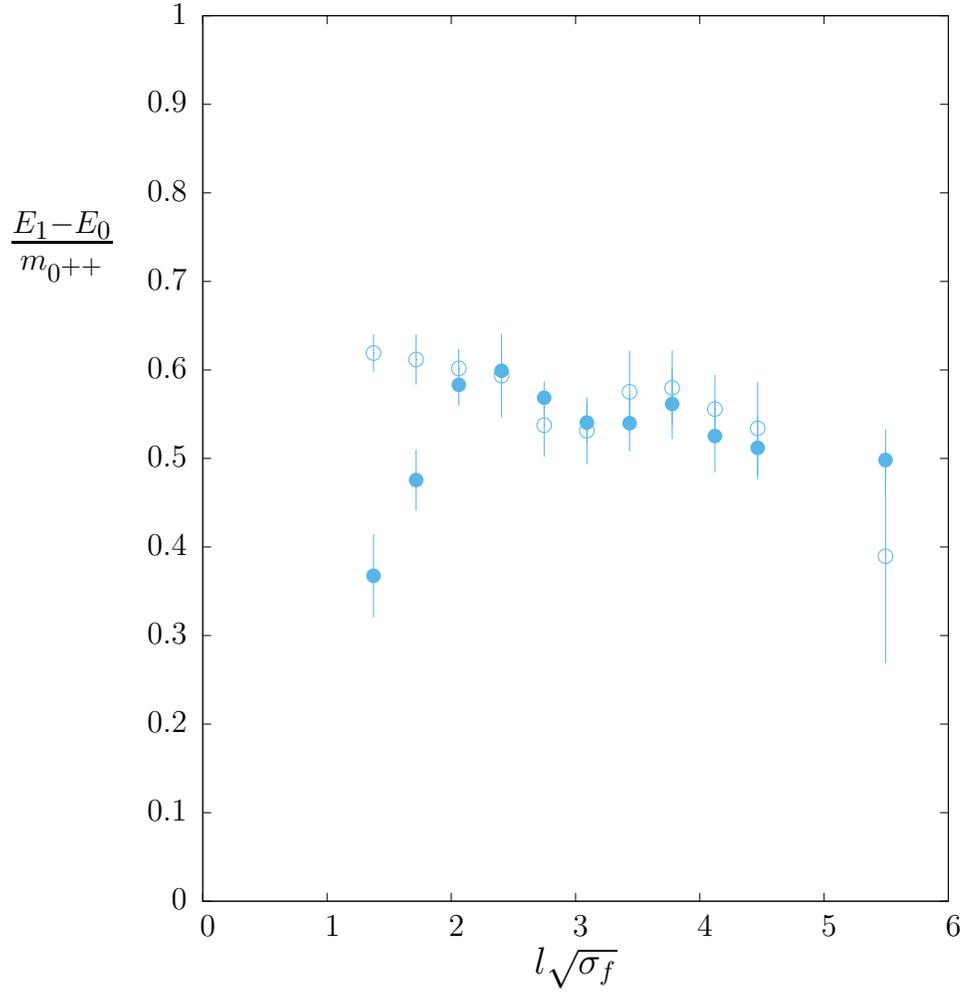}
\end	{center}
\caption{ Difference between ground and first excited energies of a $k=2A$ , $\bullet$, 
  and $k=3A$, $\circ$, flux tubes of length $l$, in units of the $l=\infty$ mass gap.
  For $SU(6)$ at $\beta=171$.}
\label{fig_DE1E0k2Ak3A_m0pp_n6}
\end{figure}

\

%
%


\begin{figure}[htb]
\begin	{center}
\input	{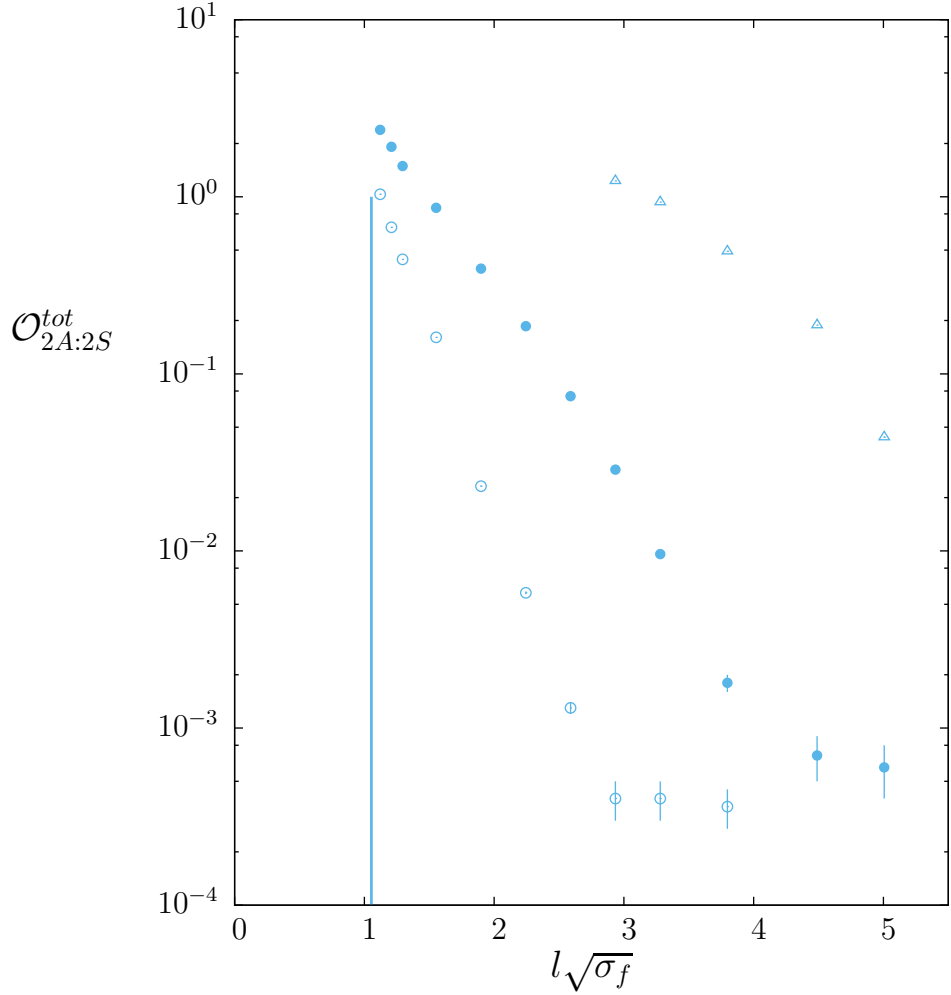}
\end	{center}
\caption{ Overlap squared of k=2A basis onto the k=2S basis, using blocking levels 1 to 4, $\circ$, 1 to 5, $\bullet$,and 1 to 6, $\vartriangle$.  Normalisation such that maximum possible overlap is 68, 85, 102  respectively. In SU(4) at $\beta=74$. Vertical line locates the deconfining transition. }
\label{fig_k2Atotk2S_n4}
\end{figure}

\begin{figure}[htb]
\begin	{center}
\input	{plot_k2Agsk2S_n4.tex}
\end	{center}
\caption{ Overlap squared of k=2A (variational) ground state onto the whole k=2S operator basis, using blocking levels 1 to 4, $\circ$, 1 to 5, $\bullet$,and 1 to 6, $\vartriangle$.  In SU(4) at $\beta=74$. Vertical line locates the deconfining transition. }
\label{fig_k2Agsk2S_n4}
\end{figure}

\begin{figure}[htb]
\begin	{center}
\input	{plot_k2Sgsk2A_n4.tex}
\end	{center}
\caption{ Overlap squared of k=2S (variational) ground state onto the whole k=2A operator basis, using blocking levels 1 to 4, $\circ$, 1 to 5, $\bullet$,and 1 to 6, $\vartriangle$.  In SU(4) at $\beta=74$. Vertical line locates the deconfining transition. }
\label{fig_k2Sgsk2A_n4}
\end{figure}

\begin{figure}[htb]
\begin	{center}
\input	{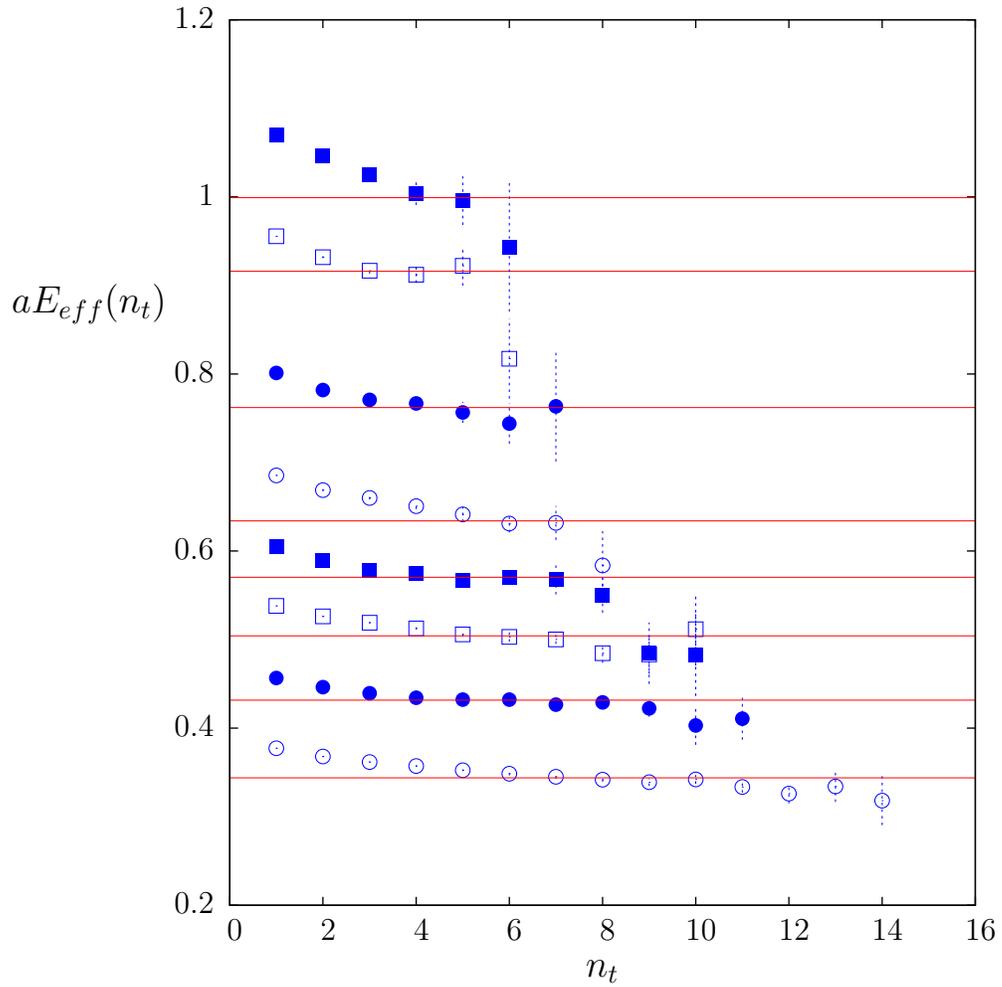}
\end	{center}
\caption{ Effective energy of the k=2S (variational) ground state versus $n_t=t/a$, for
$l=22,26,30,34,38,44,52,58$ (in ascending order), in $SU(4)$ at $\beta=74$.}
\label{fig_Eeffk2Sgs_n4}
\end{figure}

\begin{figure}[htb]
\begin	{center}
\input	{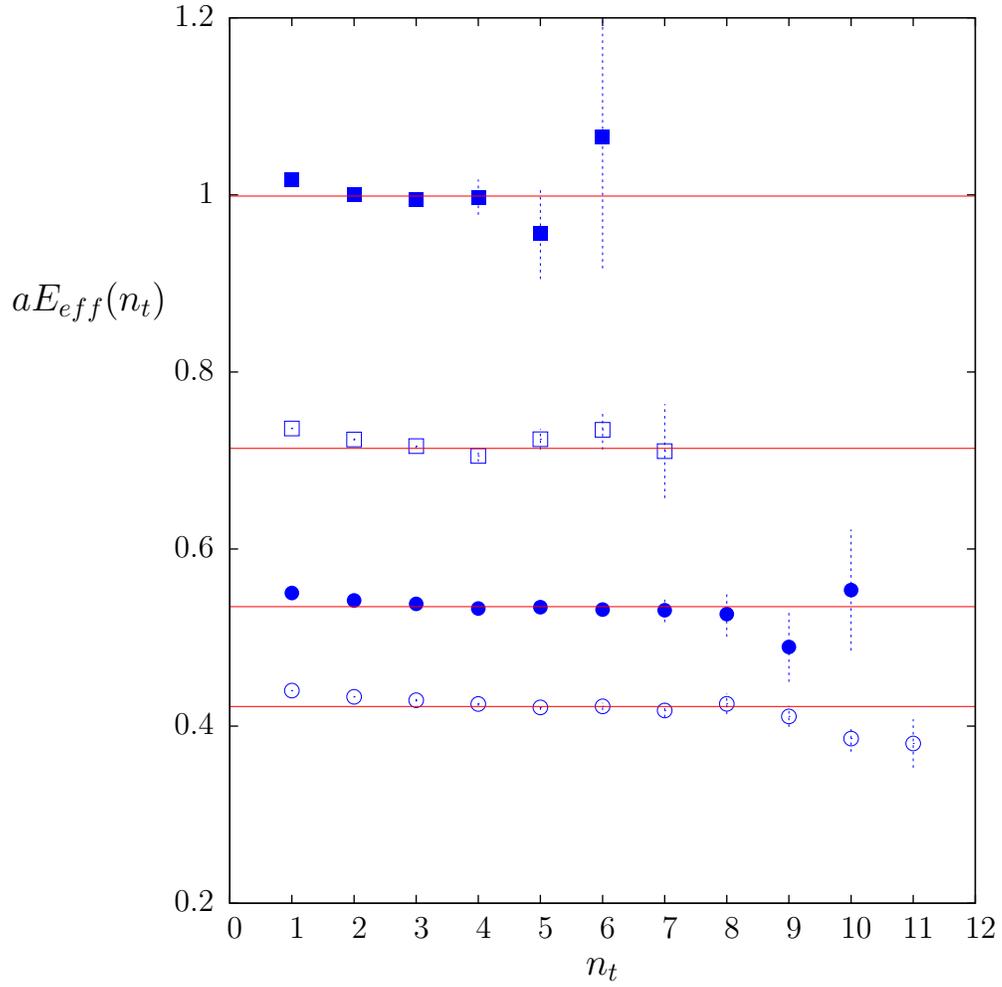}
\end	{center}
\caption{ Effective energy of the k=2S (variational) ground state versus $n_t=t/a$, for
$l=28,34,44,60$ (in ascending order), in $SU(8)$ at $\beta=306.25$.}
\label{fig_Eeffk2Sgs_n8}
\end{figure}

\clearpage


\begin{figure}[htb]
\begin	{center}
\input	{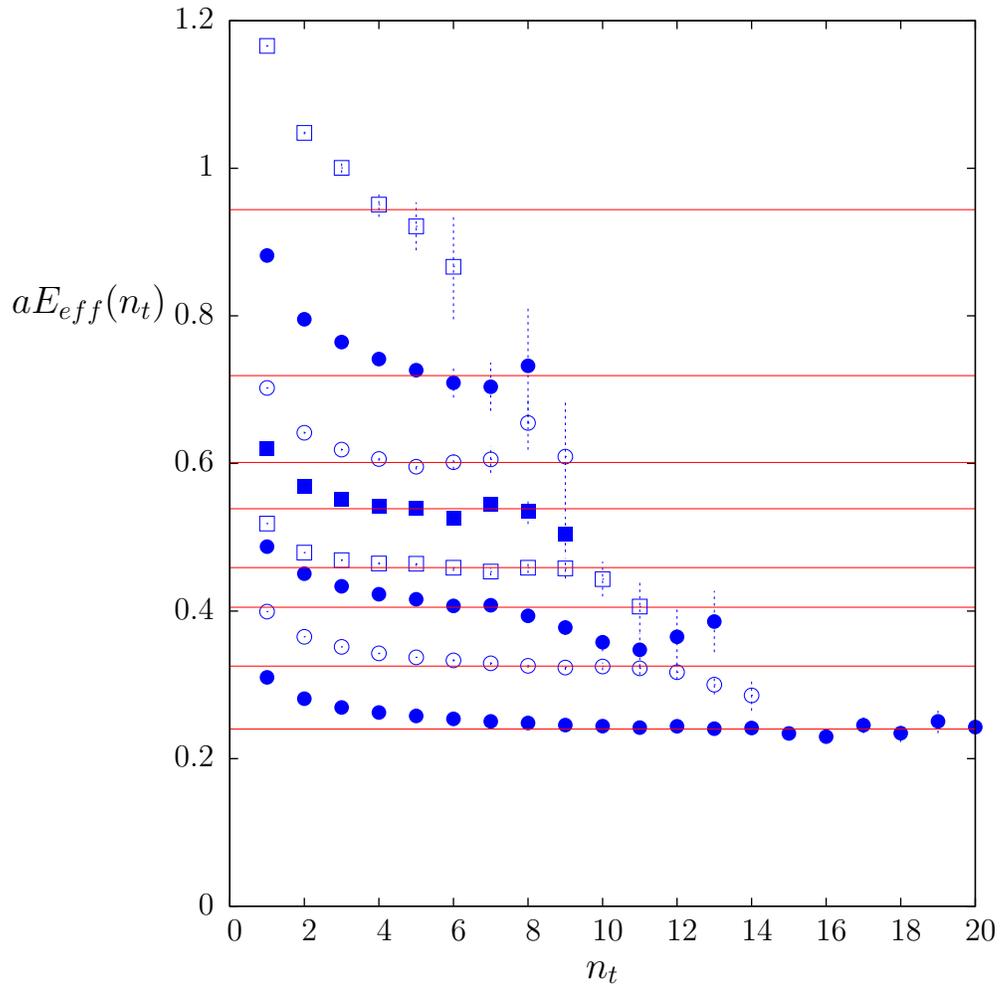}
\end	{center}
\caption{ Effective energy of the adjoint k=0 ground state versus $n_t=t/a$, for
$l=18,22,26,30,34,38,46,58$ (in ascending order), in $SU(2)$ at $\beta=16$.}
\label{fig_Eeffk0gs_n2}
\end{figure}

\begin{figure}[htb]
\begin	{center}
\input	{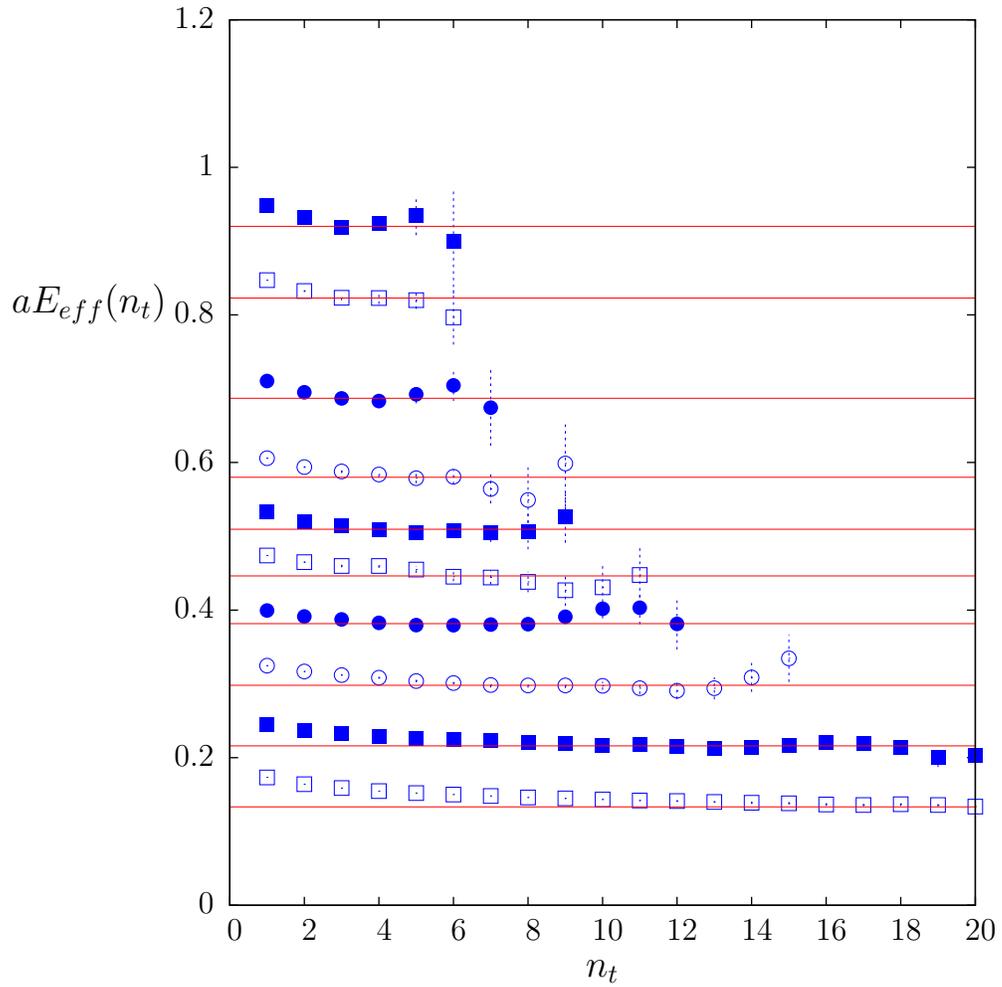}
\end	{center}
\caption{ Effective energy of the adjoint k=0 ground state versus $n_t=t/a$, for
$l=15,18,22,26,30,34,38,44,52,58$ (in ascending order), in $SU(4)$ at $\beta=74$.}
\label{fig_Eeffk0gs_n4}
\end{figure}

\begin{figure}[htb]
\begin	{center}
\input	{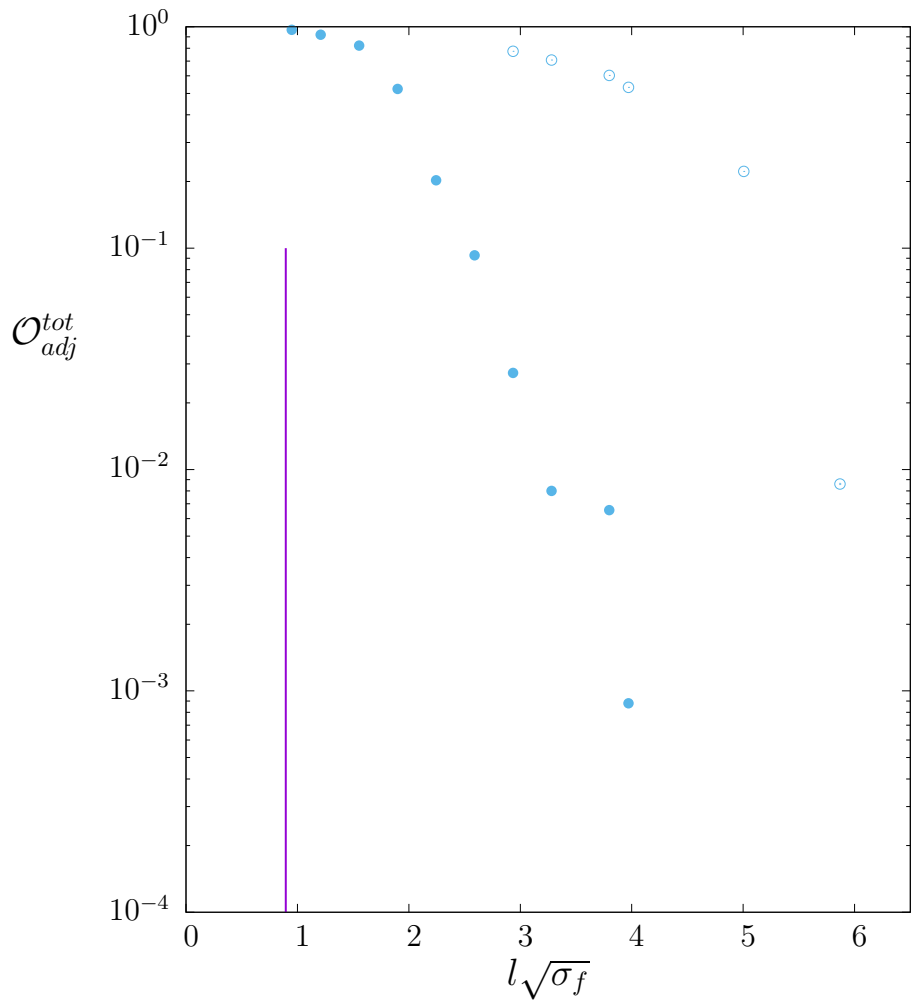}
\end	{center}
\caption{ Overlap squared of our adjoint flux loop basis on to vacuum, in
  SU(2) at $\beta=16$. Operators up to blocking level 5, $\bullet$, and up to 6, $\circ$.
  Vertical line denotes deconfining transition.}
  \label{fig_adjtotvac_n2}
\end{figure}

\begin{figure}[htb]
\begin	{center}
\input	{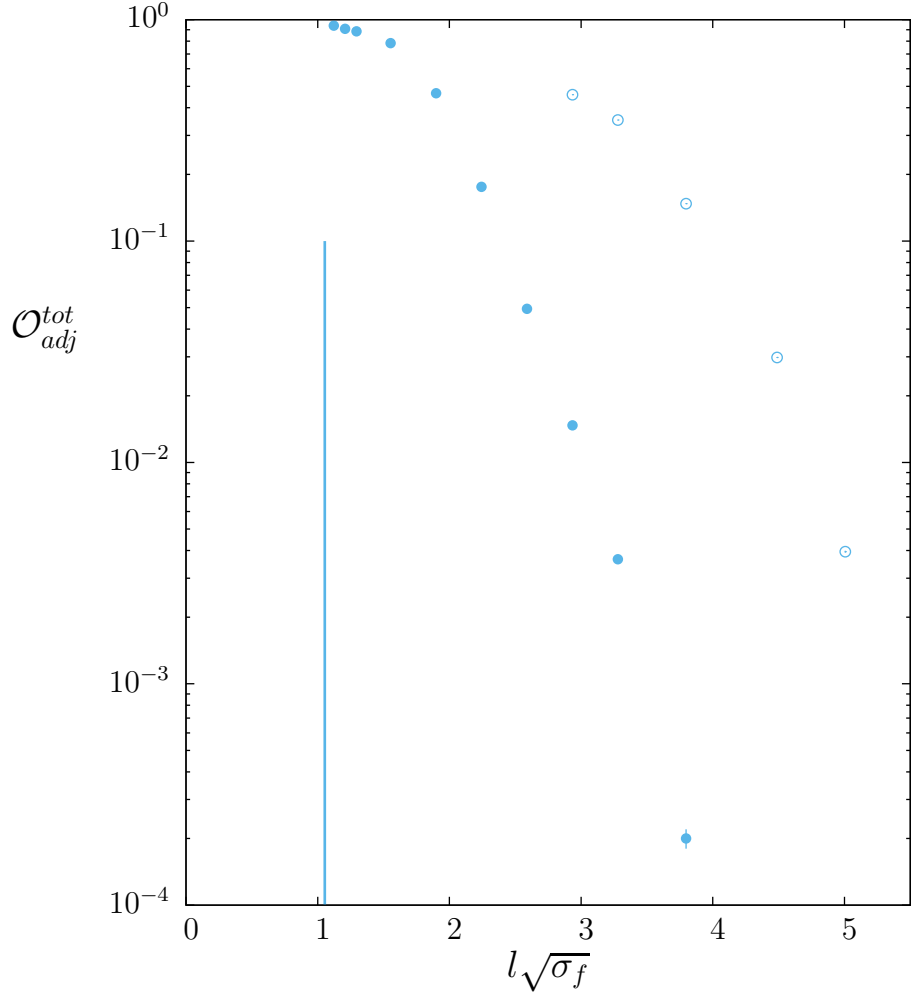}
\end	{center}
\caption{ Overlap squared of our adjoint flux loop basis on to the vacuum, 
where basis includes operators up to blocking level 5, $\bullet$, and 6, $\circ$. 
In SU(4) at $\beta=74$.  Vertical line denotes deconfining transition.}
\label{fig_adjtotvac_n4}
\end{figure}

\begin{figure}[htb]
\begin	{center}
\input	{plot_adjgsvac_n4.tex}
\end	{center}
\caption{ Overlap squared of the best variational ground state adjoint flux 
loop operator onto to vacuum, where basis includes operators up to blocking 
level 5, $\bullet$, and 6, $\circ$. In SU(4) at $\beta=74$.
  Vertical line denotes deconfining transition.}
\label{fig_adjgsvac_n4}
\end{figure}

\begin{figure}[htb]
\begin	{center}
\input	{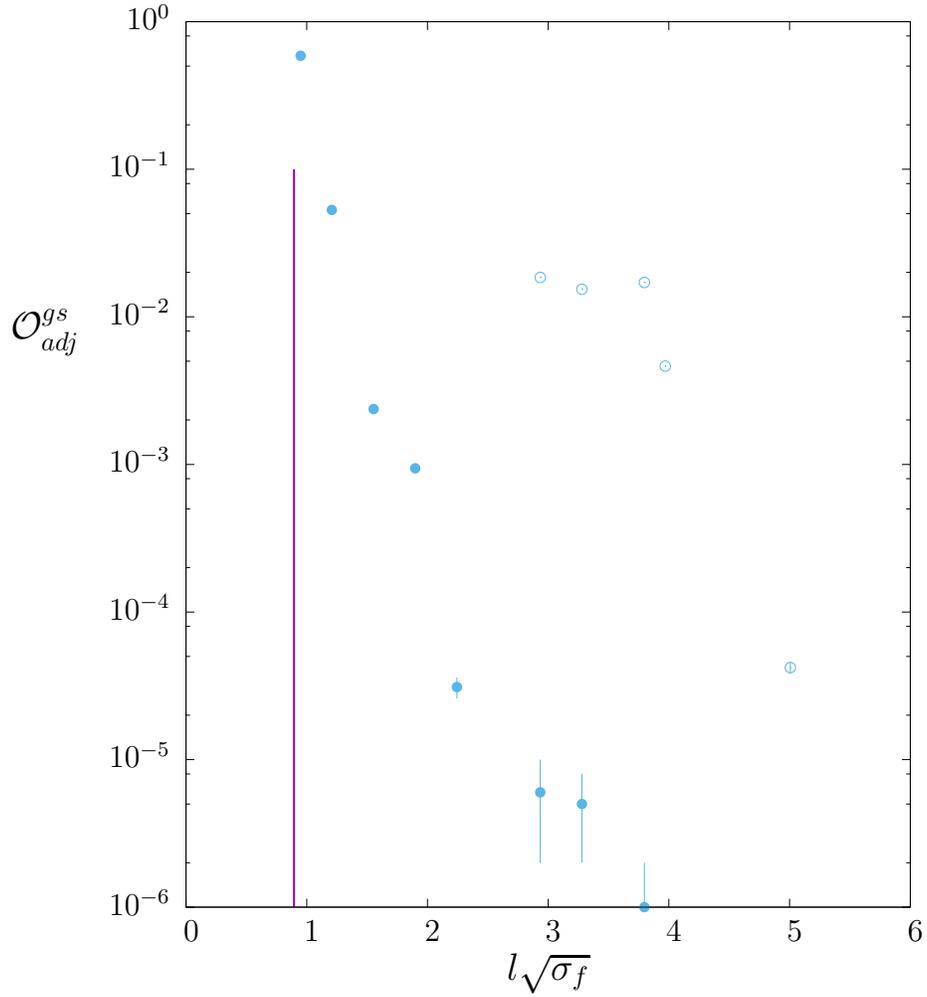}
\end	{center}
\caption{ Overlap squared of the best ground state adjoint flux loop operator onto to vacuum, in
SU(2) at $\beta=16$. Operators up to blocking level 5, $\bullet$, and up to 6, $\circ$.
  Vertical line denotes deconfining transition.}
\label{fig_adjgsvac_n2}
\end{figure}

%
%


\begin{figure}[htb]
\begin	{center}
\input	{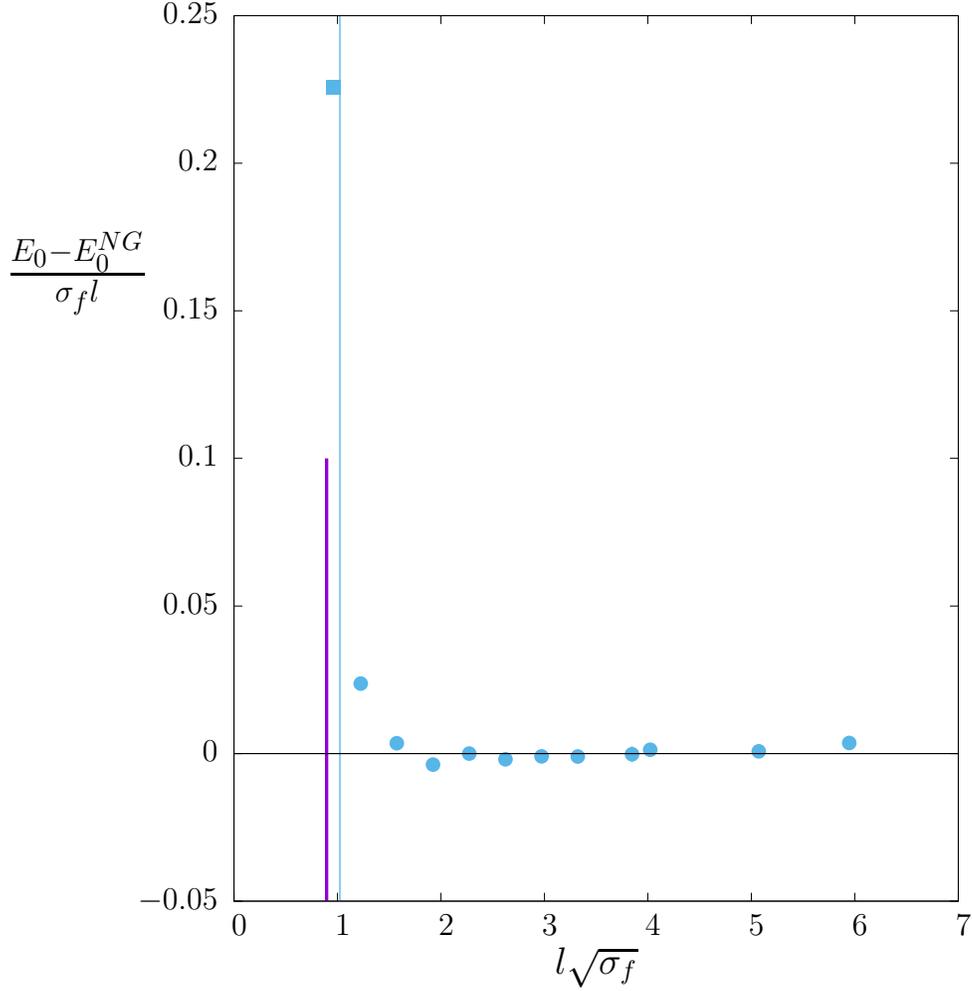}
\end	{center}
\caption{Ground state energy minus Nambu-Goto energy in $SU(2)$ at $\beta=16$.
  To left of light vertical line $E_0^{NG}$ is tachyonic and for $\blacksquare$
  we have set it to zero. Thick vertical line locates the deconfining transition. }
\label{fig_DENG_gs_su2}
\end{figure}

\begin{figure}[htb]
\begin	{center}
\input	{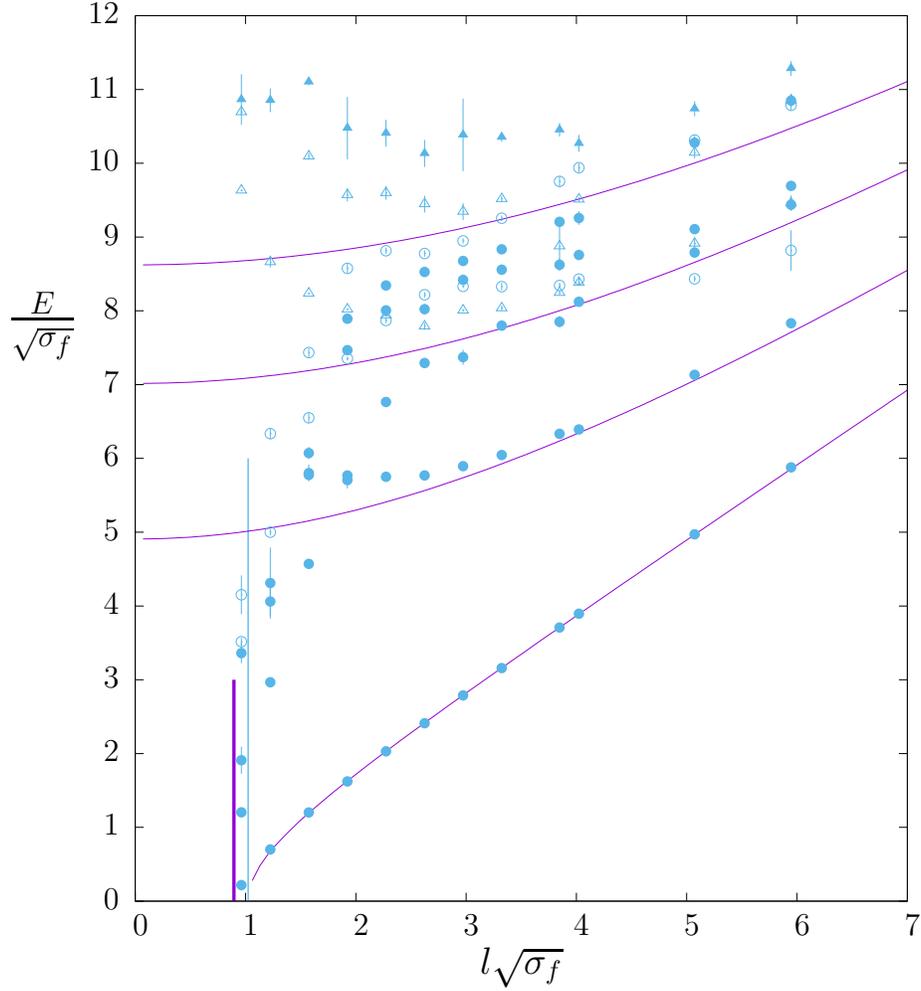}
\end	{center}
\caption{SU(2) at $\beta=16$. Spectrum with $p=0$ and parities $P_t,P_{\parallel}=+,+$,
  $\bullet$, $-,+$, $\circ$, $+,-$, $\blacktriangle$, $-,-$, $\vartriangle$. Thick
  vertical line is deconfining transition; thin vertical is NG tachyonic transition.}
\label{fig_Ek1_q0Pall_su2}
\end{figure}

\begin{figure}[htb]
\begin	{center}
\input	{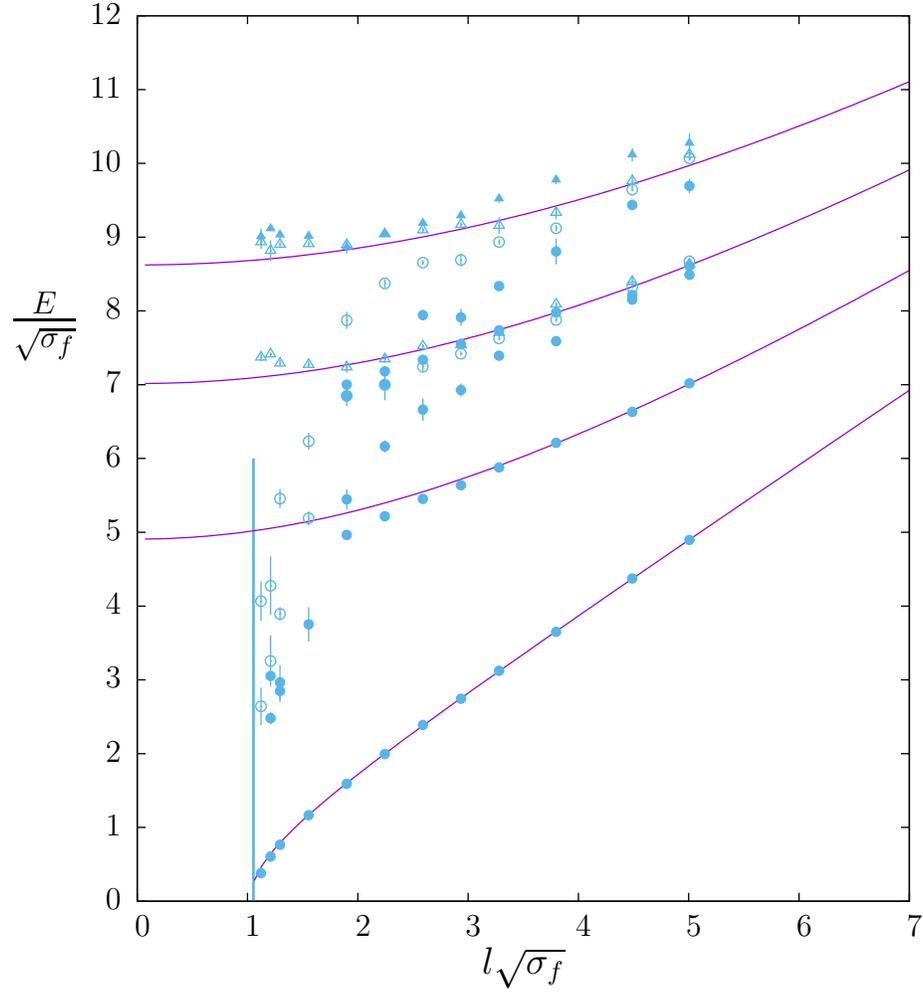}
\end	{center}
\caption{SU(4) at $\beta=74$. Spectrum with $p=0$ and parities $P_t,P_{\parallel}=+,+$,
  $\bullet$, $-,+$, $\circ$, $+,-$, $\blacktriangle$, $-,-$, $\vartriangle$. 
  Vertical line is deconfining transition.}
\label{fig_Ek1_q0Pall_su4}
\end{figure}


\begin{figure}[htb]
\begin	{center}
\input	{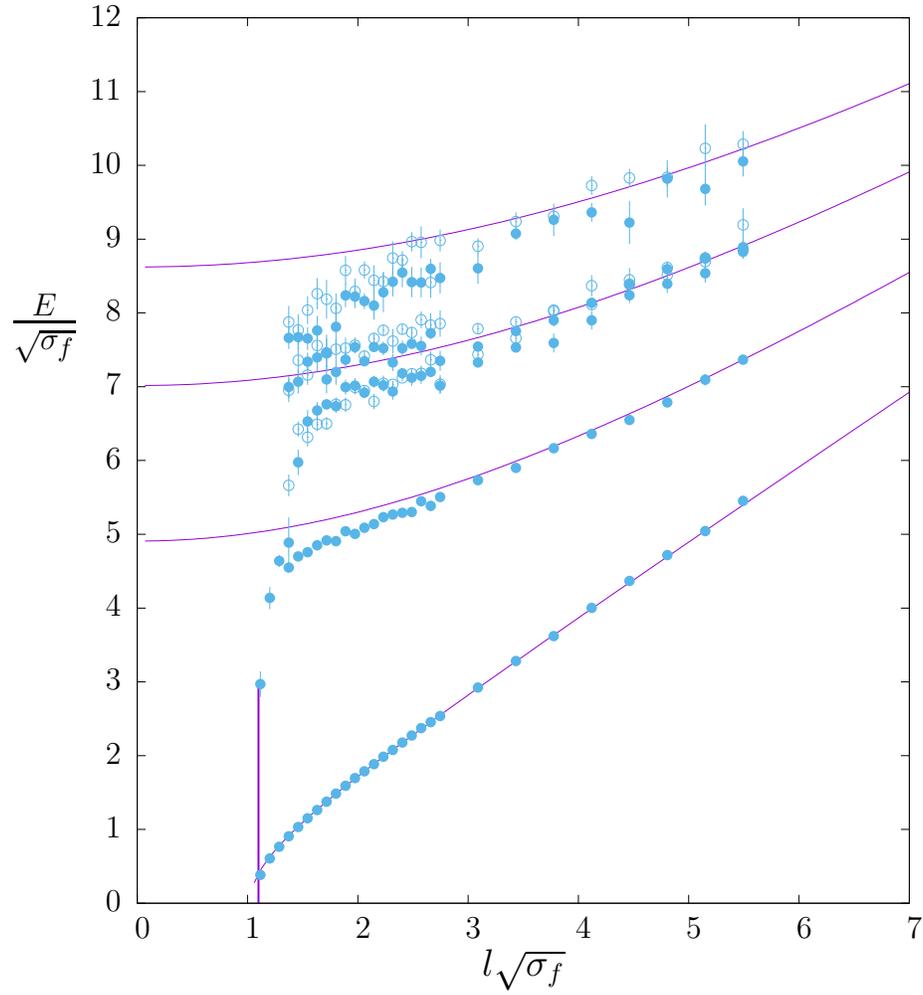}
\end	{center}
\caption{SU(6) at $\beta=171$. Spectrum with $p=0$ and parities $P_t=+$,
  $\bullet$, and  $P_t=-$, $\circ$. Vertical line is deconfining transition}
\label{fig_Ek1_q0Pall_su6}
\end{figure}

\begin{figure}[htb]
\begin	{center}
\input	{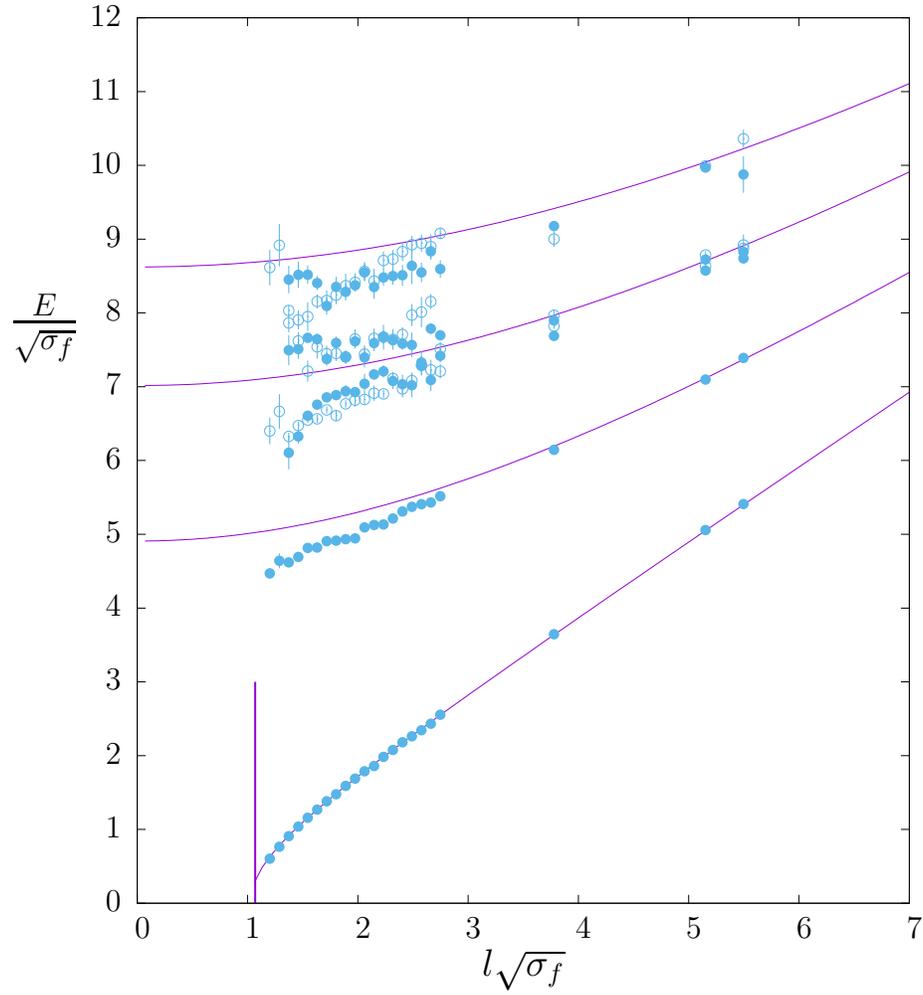}
\end	{center}
\caption{SU(8) at $\beta=306.25$. Spectrum with $p=0$ and parities $P_t=+$,
  $\bullet$, and  $P_t=-$, $\circ$. Values at $l/a=44,60$ are from
  \cite{AAMT16}. Vertical line is deconfining transition}
\label{fig_Ek1_q0Pall_su8}
\end{figure}

\clearpage


\begin{figure}[htb]
\begin	{center}
\input	{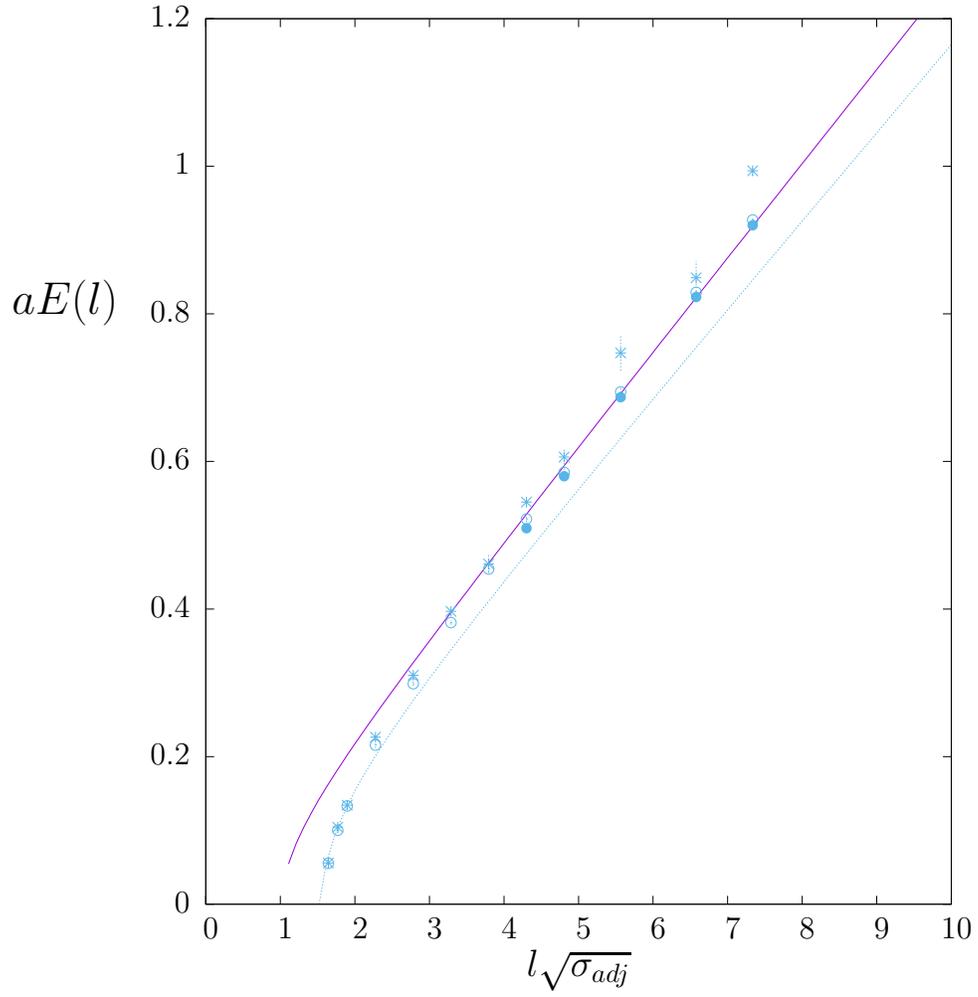}
\end	{center}
\caption{Adjoint flux tube ground state energies for SU(4) at $\beta=74$
  for operator bases with blocking levels $\leq 4$
  ($\times$), $\leq 5$ ($\circ$), and $\leq 6$
  ($\bullet$). Solid line is Nambu-Goto fit to largest blocking levels;
dashed line is twice the energy of a fundamental flux tube.}
\label{fig_Egsk0bl456b_su4}
\end{figure}

\begin{figure}[htb]
\begin	{center}
\input	{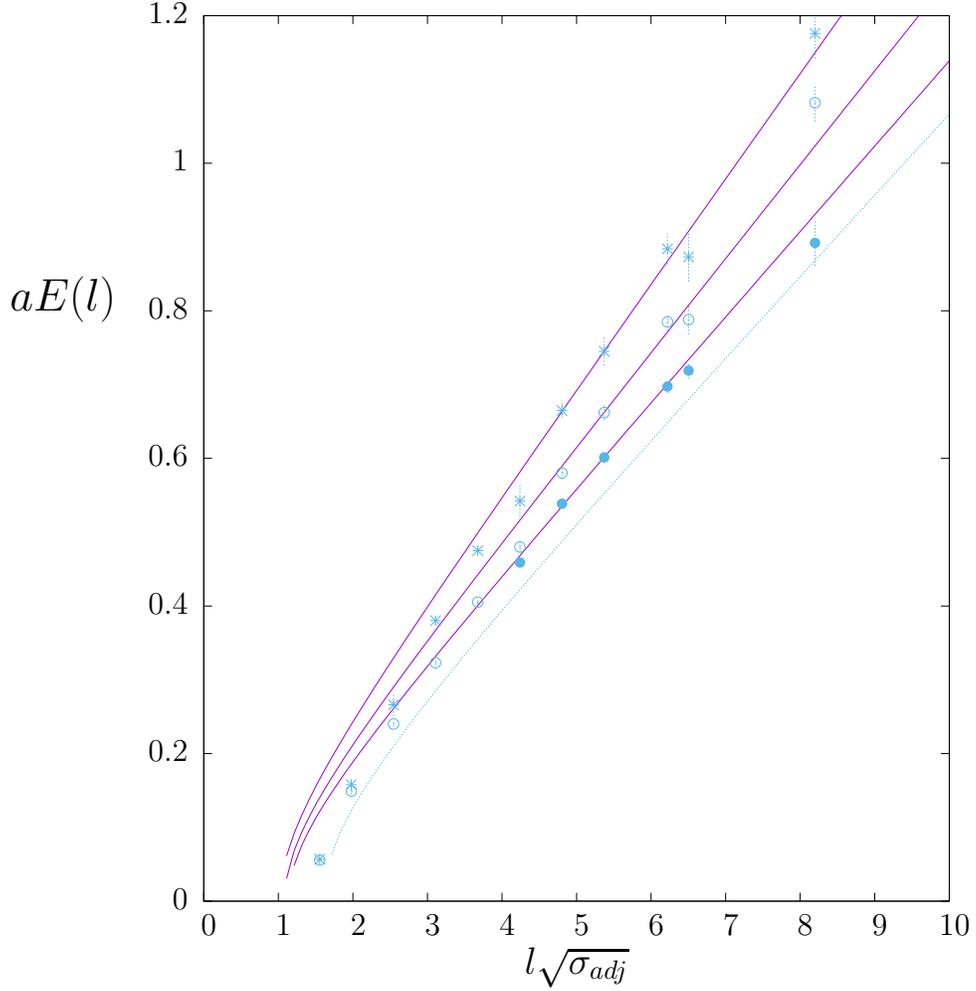}
\end	{center}
\caption{Adjoint flux tube ground state energies for SU(2) at $\beta=16$
  for operator bases with blocking levels $\leq 4$
  ($\times$), $\leq 5$ ($\circ$), and $\leq 6$
  ($\bullet$). Solid lines are Nambu-Goto fits to largest $l$;
dashed line is twice the energy of a fundamental flux tube.}
\label{fig_Egsk0bl456b_su2}
\end{figure}

\begin{figure}[htb]
\begin	{center}
\input	{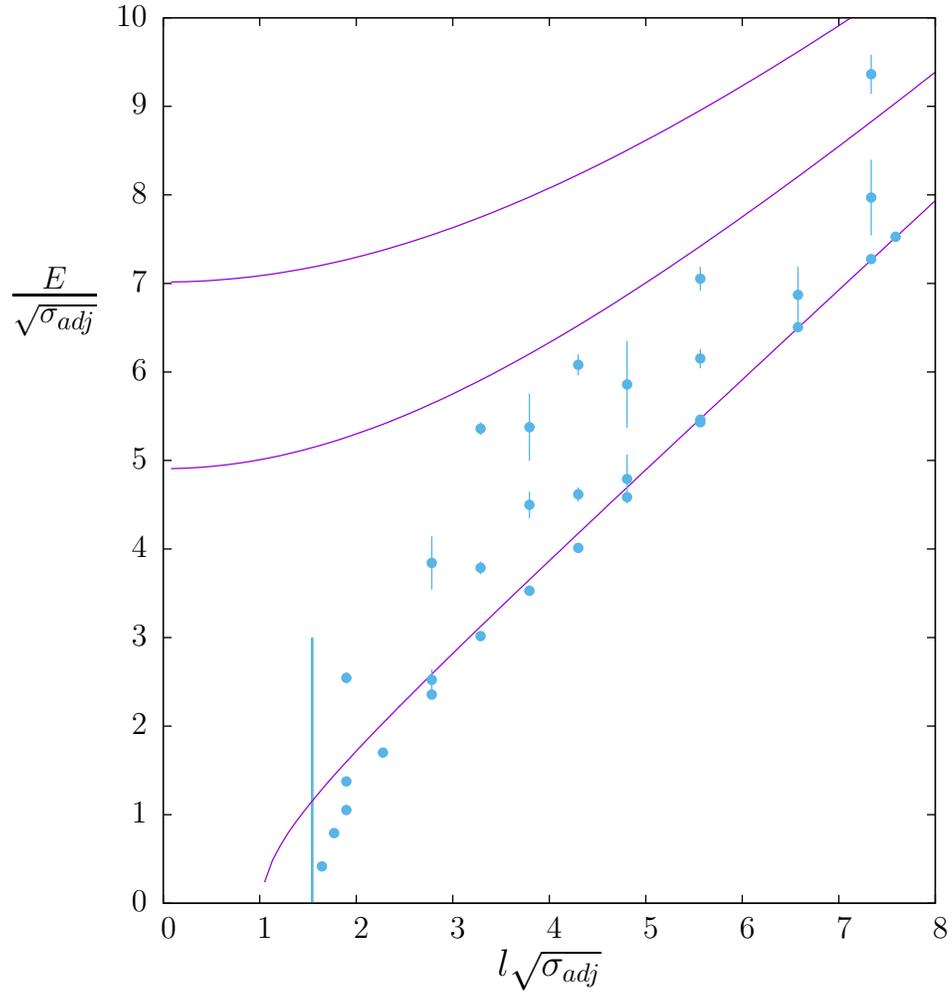}
\end	{center}
\caption{Adjoint flux tube in SU(4) at $\beta=74$. Spectrum with $p=0$ and
  parities $P_t,P_{\parallel}=+,+$.
Vertical line locates deconfining transition.}
\label{fig_Ek0_q0P++_su4}
\end{figure}


\begin{figure}[htb]
\begin	{center}
\input	{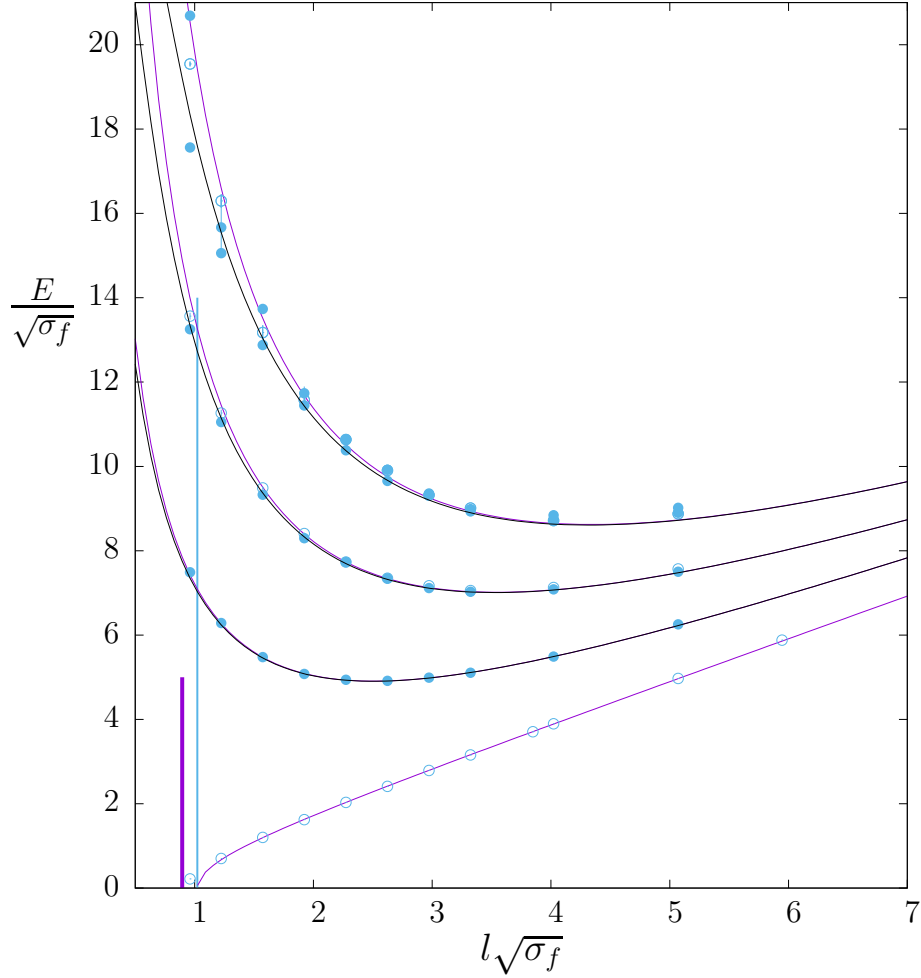}
\end	{center}
\caption{SU(2) at $\beta=16$. Ground state energies for momenta
  $p=0,2\pi/l,4\pi/l,6\pi/l$ and parities $P=+$, $\bullet$, and
  $P=-$, $\circ$. Solid curves are continuum NG; dashed curves are NG
  with a 'lattice' dispersion relation. Thick vertical line is deconfining transition;
  thin vertical line is NG tachyonic transition.}
\label{fig_E_gsqall_su2}
\end{figure}

\begin{figure}[htb]
\begin	{center}
\input	{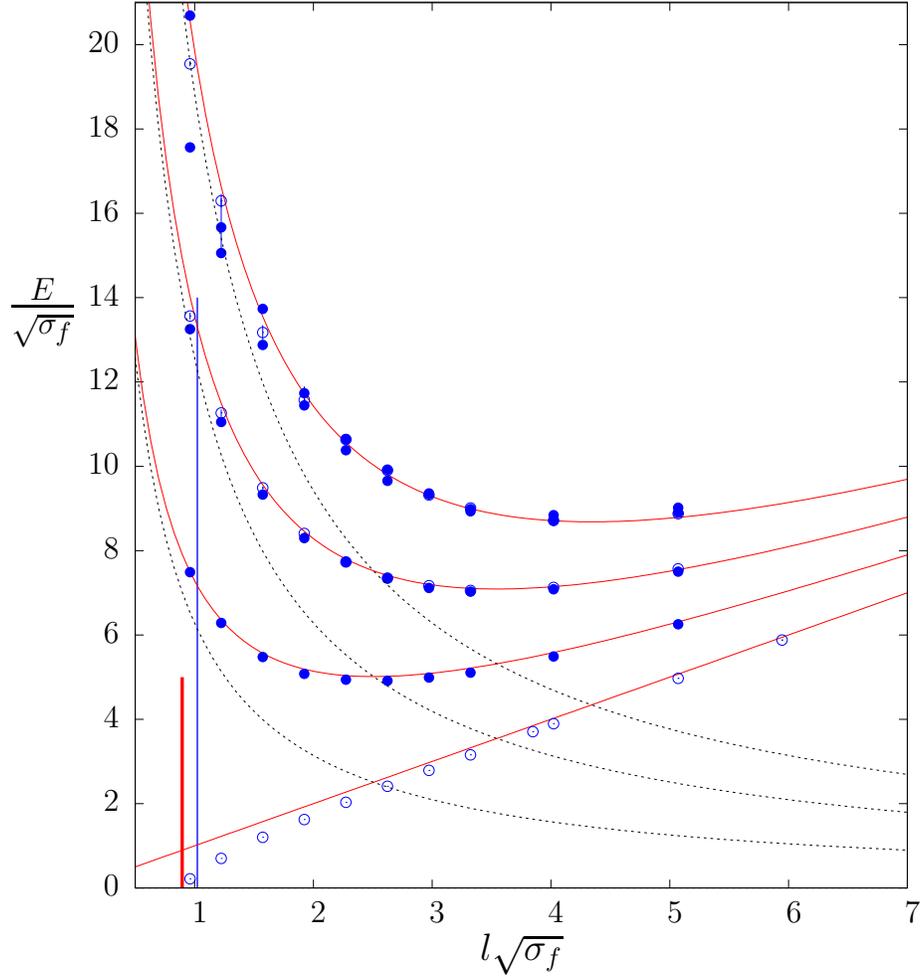}
\end	{center}
\caption{SU(2) at $\beta=16$. Ground state energies for momenta
  $p=0,2\pi/l,4\pi/l,6\pi/l$ and parities $P=+$, $\bullet$, and
  $P=-$, $\circ$. Solid curves are
  the prediction for free phonons on a background string, 
  $E=l\sigma_f + p$. Dashed curves are
  the continuum prediction for just a massless free particle,
  $E=p=2\pi q/l$. Thick vertical line is deconfining transition;
  thin vertical line is NG tachyonic transition.}
\label{fig_E_gsqall_fits_su2}
\end{figure}

\begin{figure}[htb]
\begin	{center}
\input	{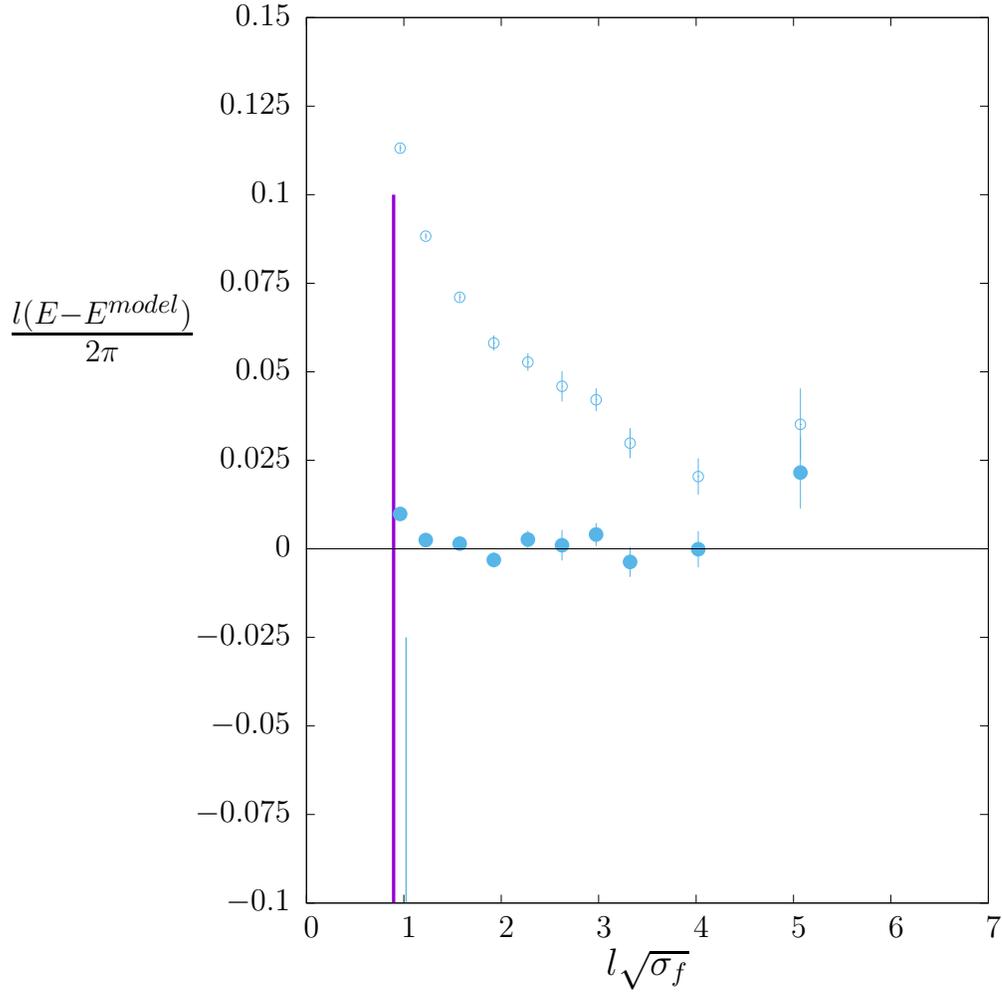}
\end	{center}
\caption{SU(2) at $\beta=16$. Energy of ground states for momenta
  $p=2\pi/l$ and parity  $P=-$ with two different model energies subtracted:
  Nambu-Goto ($\bullet$) and free massles phonons ($\circ$) on a
  background flux tube.
  Thick vertical line is deconfining transition; thin vertical line is
  Nambu-Goto tachyonic transition.}
\label{fig_DEmodelgs_k1q1P-_su2}
\end{figure}

\begin{figure}[htb]
\begin	{center}
\input	{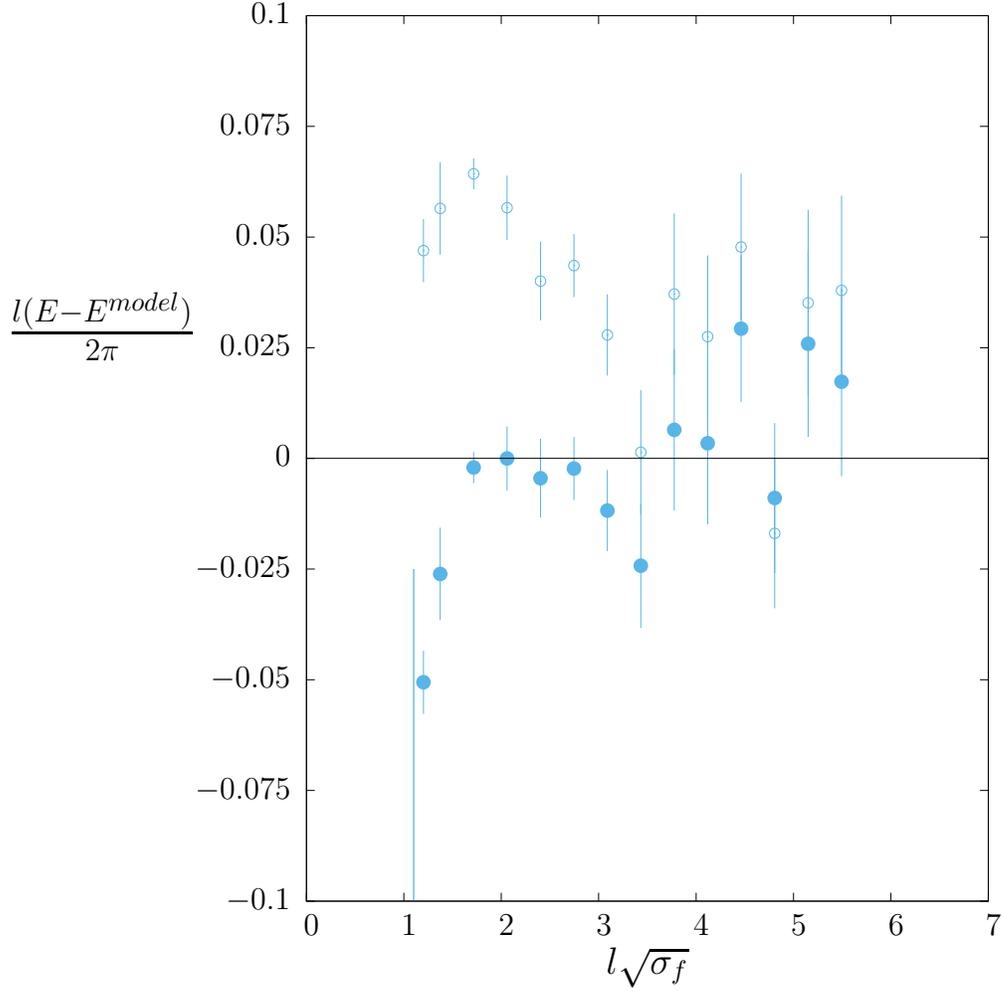}
\end	{center}
\caption{SU(6) at $\beta=171$. Energy of ground states for momenta
  $p=2\pi/l$ and parity  $P=-$ with two different model energies subtracted:
  Nambu-Goto ($\bullet$) and free massles phonons ($\circ$) on a
  background flux tube.
  Vertical line is deconfining transition.}
\label{fig_DEmodelgs_k1q1P-_su6}
\end{figure}

\begin{figure}[htb]
\begin	{center}
\input	{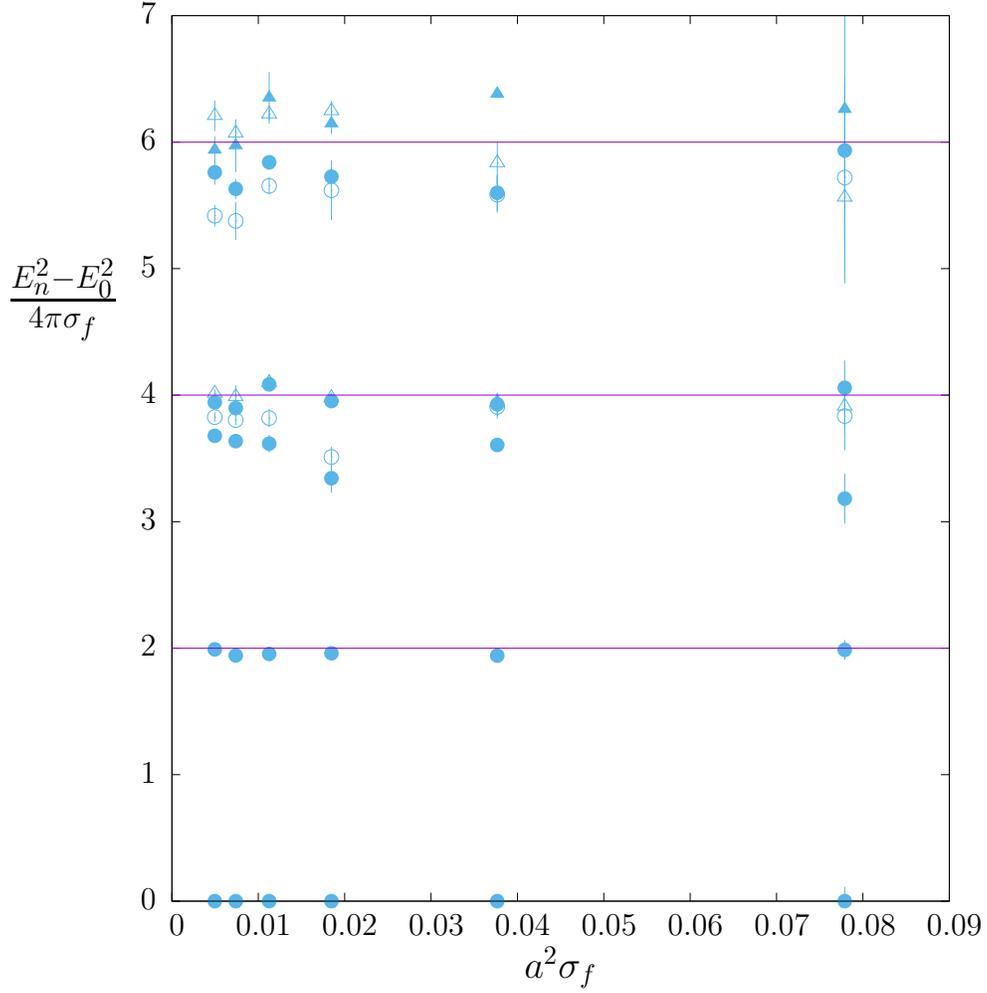}
\end	{center}
\caption{Approach to continuum limit in SU(8). Flus tube lengths 
  lie in narrow range $l\surd\sigma_f \in (3.78,3.91)$.  Spectrum with $p=0$
  and parities
  $P_t,P_{\parallel}=+,+$, $\bullet$, $-,+$, $\circ$, $+,-$, $\blacktriangle$, $-,-$,
  $\vartriangle$.} 
\label{fig_DENGk1_cont_n8}
\end{figure}

\begin{figure}[htb]
\begin	{center}
\input	{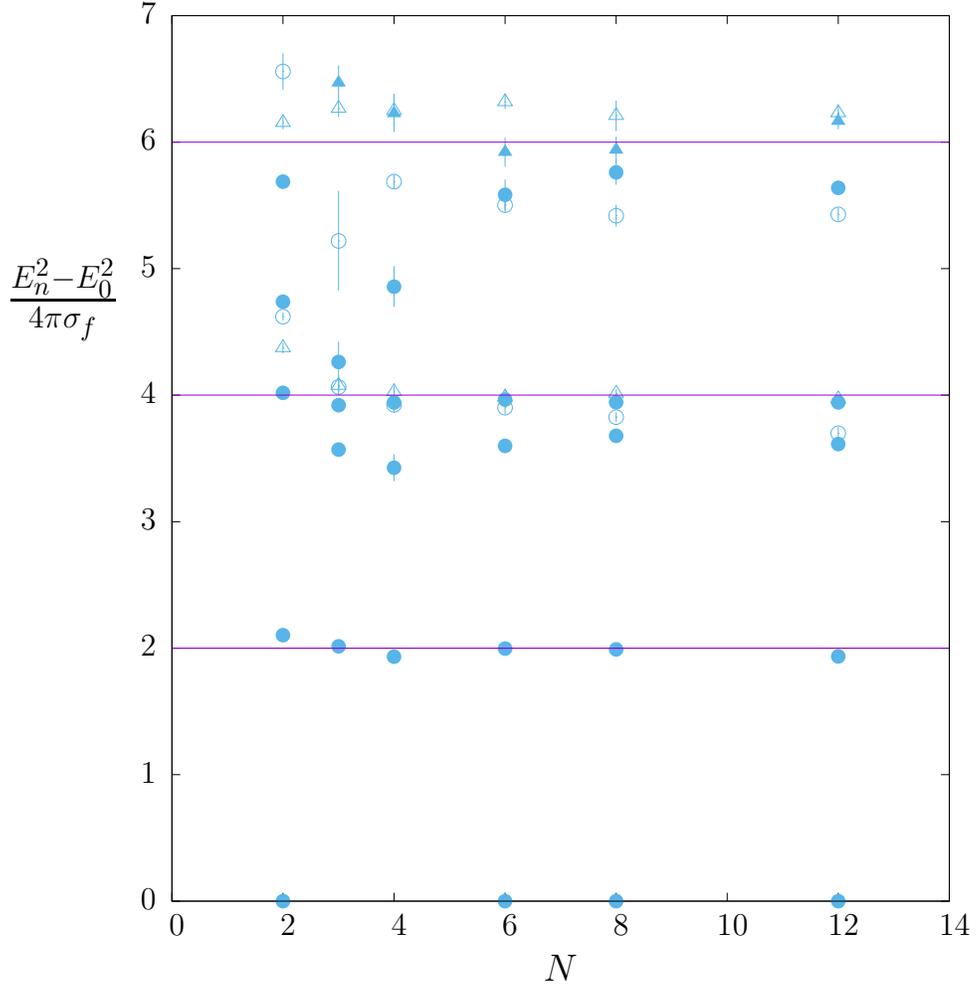}
\end	{center}
\caption{Variation of  $p=0$ flux tube spectrum with $N$. Lengths and lattice spacings
  lie in narrow ranges $l\surd\sigma_f \in (3.55,3.85)$ and $a\surd\sigma_f \in (0.069,0.074)$.
  Parities  $P_t,P_{\parallel}=+,+$, $\bullet$, $-,+$, $\circ$, $+,-$, $\blacktriangle$,
  $-,-$, $\vartriangle$.} 
\label{fig_DENGk1_suNc}
\end{figure}

\end{document}